\documentclass[aps,floatfix,twocolumn,prl,superscriptaddress,10pt]{revtex4-2}
\usepackage{blindtext,bm,amsmath,amsfonts,physics,tikz,hyperref,siunitx,mathtools}
\usetikzlibrary{decorations.pathmorphing,decorations.pathreplacing,decorations.markings,arrows.meta,shapes.geometric,calc,fadings,shadings,bending}
\newcommand{\lo}[1]{_{#1}}
\newcommand{\hi}[1]{^{#1}}
\newcommand{\mn}{\mu\nu}
\newcommand{\dm}[1]{\partial_{#1}}
\newcommand{\ud}{\mathrm{d}}

\newcommand{\pk}[1]{\psi_{\bm{#1}}}
\newcommand{\dg}[0]{^{\dagger}}
\newcommand{\pkd}[1]{\psi\dg_{\bm{#1}}}

\definecolor{chalmershimmel}{RGB}{88,176,227}
\definecolor{chalmerskoppar}{RGB}{0,169,157}
\definecolor{chalmerstegel}{RGB}{241,90,34}
\definecolor{chalmersvastkust}{RGB}{0,48,80}
\definecolor{chalmersenergi}{RGB}{255, 203, 5}
\DeclareSIUnit\angstrom{\text {Å}}
\usepackage{xcolor}

\hypersetup{
  colorlinks,
  citecolor=chalmersvastkust,
  linkcolor=chalmersvastkust,
  urlcolor=chalmersvastkust}

\begin{document}

%\preprint{}

\title{Inertial Repulsion from Quantum Geometry}


\author{Maike Fahrensohn}
%\thanks{}
\affiliation{Department of Physics, Chalmers University of Technology, 412 96 Göteborg, Sweden}
\author{R. Matthias Geilhufe}
\affiliation{Department of Physics, Chalmers University of Technology, 412 96 Göteborg, Sweden}

\date{\today}

\begin{abstract}
%... by treating as a full gauge theory...
We derive a repulsive, charge-dipole-like interaction for a Dirac particle in a rotating frame, arising from a geometric $U(1)$ gauge symmetry associated with the Berry phase. The Lagrangian of this system includes a non-inertial correction due to centrifugal field coupling. By imposing gauge symmetry and treating it as a full gauge theory, the Lagrangian is extended to include Berry connection and curvature terms. Upon integrating out the geometric gauge field, the effective action is obtained. This leads to the emergence of a repulsive, long-range effective interaction in the Lagrangian. Explicitly, in the non-inertial frame of the observer, the geometric gauge invariance effectively leads to a repulsive Coulomb-interaction in momentum space. In real space, the inertial repulsion manifests in a $1/\vert r\vert\hi{2}$ potential, which is symmetric about the origin of rotation and mirrors charge-dipole interaction.

\end{abstract}


\maketitle

%\section{\label{sec:intro}Introduction}

In classical mechanics, inertial forces are the fictitious forces that arise in non-inertial frames of reference due to acceleration and rotation. These include the Coriolis force, the centrifugal force, and the Euler force ~\cite{demtroder2017mechanics,corben2013classical}.
 %As shown by Hehl and Ni \cite{hehl1990inertial} and further developed by Geilhufe \cite{geilhufe2022dynamic}, 
Inertial effects extend naturally to quantum mechanical systems~\cite{hehl1990inertial,Matsuo2011,Matsuo2011prb,geilhufe2022dynamic}. %Quantum systems described in non-inertial frames, such as the rest frame of an electron bound to a rotating ion, the system's Hamiltonian acquires additional inertial terms. These include spin-rotation coupling, centrifugal field coupling, centrifugal spin-orbit interaction, and the emergence of a centrifugal redshift.
%It is worth noting that such non-inertial systems are not confined to rotating laboratory systems. As Hehl and Ni pointed out \cite{hehl1990inertial}, 
For example, %Even 
stationary laboratories on Earth constitute non-inertial systems as they accelerate and rotate relative to local inertial frames~\cite{thorne2000gravitation,ni1978inertial,ni1978inertial,defacio1978presymmetry}. Their influence on the quantum mechanical wave function has been verified using neutron interferometry~\cite{Page1975,mashhoon1988neutron,Werner1979}. 
Excitingly, the concept of inertial effects also holds on the nanoscale and has been discussed in the context of nano-resonators~\cite{Matsuo2011,Matsuo2011prb} or axial phonons and molecular rotations~\cite{geilhufe2022dynamic,Huang2024}. Hence, any accelerating and rotating system will, in its co-moving frame, experience similar corrections to the Hamiltonian. These ideas are, therefore, applicable to a wide variety of physical systems. %Furthermore, the mechanism presented in this paper is expected to extend to other systems exhibiting interaction terms coupling linearly to the position operator that are not inertial in nature. 

In this paper, we focus on the effect of the centrifugal force on quantum particles in a rotating frame and show that its connection to quantum geometry induces a repulsive effective force. As the centrifugal force is odd under spatial inversion, it couples linearly to the position operator of the electron \cite{geilhufe2022dynamic,hehl1990inertial}, similarly to the electric dipole interaction. In momentum space, this coupling shows up as a derivative which can be promoted to a covariant derivative under a local $U(1)$ geometric gauge transformation associated with the Berry phase \cite{pancharatnam1956generalized,berry1984quantal}. 

The use of gauge fields plays an important role in modern physics, permeating a wide range of fields from high energy and nuclear physics to condensed matter \cite{peskin2018introduction,pokorski2000gauge}. Inter alia, they mediate fundamental interactions \cite{mills1989gauge}, describe gravitation \cite{ivanenko1983gauge,ryder2009introduction}, and emerge as effective descriptions of geometric phases \cite{ivanenko1983gauge,o1997dawning,o2000gauge}. In adiabatic quantum systems, the Berry connection acts as a geometric gauge field \cite{fujita2011gauge,palumbo2019tensor}. In the context of the dipole coupling to the centrifugal force, we treat the Berry curvature as a dynamical degree of freedom of the theory that, upon integration, can mediate effective forces and leads to a charge-dipole-like repulsive effective interaction, emerging from the geometry of quantum phase space. 

%In section \ref{sec:model}, we introduce the model Hamiltonian with the centrifugal field coupling term and the geometric gauge field. Section \ref{sec:inertialrep} derives the inertial repulsion from the model Lagrangian by integrating out the geometric gauge field and obtaining a repulsive potential in real space. The results and possible extensions of this work are discussed in section \ref{sec:discussion}.
%\section{\label{sec:model}Model and Geometric Coupling}
% \begin{itemize}
%     \item introduce Lagrangian
%     \item argue why we can add Berry curvature square term
%     \item talk about geometric U(1) gauge symmetry and how taking this to be the Berry phase leads to the Berry potential and the Berry curvature
% \end{itemize}
In the following, we develop the theory for a Dirac particle in a rotating frame. However, it should be noted, that the effect is more general and can be extended to Dirac particles in accelerated systems.
%In the following, we develop the theory on the example of the electron, as the paradigmatic representative of a fermion. However, we note that the effect is more general and holds for other particles in a similar way. 

% \begin{figure}
%     \centering
%     \includegraphics[width=0.5\linewidth]{schematic-crop}
%     \caption{\textcolor{red}{[TRY TO BRING IN THE PLOT ILLUSTRATING THE EFFECT]}}
%     \label{fig:placeholder}
% \end{figure}
%An electron bound to an ion undergoing rotational motion is naturally described by the Dirac equation \cite{peskin2018introduction}. Transforming into 
In the observer's local non-inertial frame, the partial derivative in the Dirac equation is promoted to a covariant derivative involving spin connection terms determined by the tetrad coordinates of the non-inertial frame \cite{hehl1990inertial,ni1978inertial}. This yields a modified Dirac equation for the rotating frame, %that can be rewritten as a Schrödinger-like equation 
with Hamiltonian \cite{geilhufe2022dynamic,hehl1990inertial}:
\begin{multline}
    H=\beta mc\hi{2}+c\bm{\alpha}\cdot\bm{p}-\bm{\omega}\cdot \bm{J}\\
    -\beta\gamma\hi{2}\bm{F}\lo{\mathrm{centr}}\cdot\bm{r}-\frac{\gamma\hi{2}}{2mc}\{\bm{F}\lo{\mathrm{centr}}\cdot\bm{r},\bm{p}\cdot\bm{\alpha}\},
    \label{eq:hamrel}
\end{multline}
where $\bm{\alpha}$ and $\beta$ denote the Dirac matrices, $\bm{p}$ and $\bm{r}$ are the momentum and position operators, $\bm{\omega}$ is the angular velocity of the rotating frame, and $\bm{J}=\bm{L}+\bm{S}$ is the total angular momentum operator given by the orbital angular momentum $\bm{L}$ and the spin $\bm{S}$. The centrifugal force is denoted as $\bm{F}\lo{\mathrm{centr}}=m\,\bm{\omega}\times\bm{\omega}\times\bm{d}$ with $m$ being the fermion mass, $\bm{\omega}$ the angular velocity of the system and $\bm{d}$ the perpendicular distance. For a more detailed derivation of the tetrad formalism and non-inertial Dirac equation \eqref{eq:hamrel}, we refer to the supplemental material and Refs. \cite{geilhufe2022dynamic} and \cite{hehl1990inertial}. 

\begin{table*}
    \centering
    \includegraphics[width=1.\textwidth]{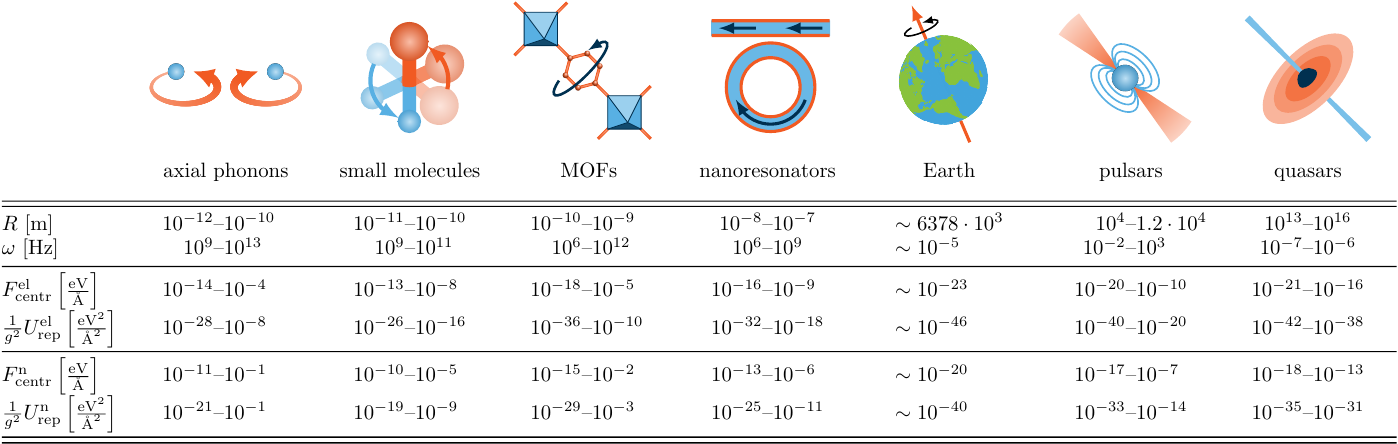}
    \caption{The centrifugal force is calculated in units of \unit{\eV/\angstrom} for a rotating electron and a rotating neutron in different physical systems. From this, the interaction strength is calculated w.r.t. the coupling constant $g$. Estimates were taken in agreement with Refs. \cite{geilhufe2022dynamic,hasegawa2005rotating,hughes2011microfabricated,alighanbari2018rotational,mudd2018quasar,hobbs2011parkes,dokuchaev2014spin,melia2001supermassive,gonzalez2019rotational,jiang2016diffusion,lovas2005diatomic,reimann2018ghz,ahn2018optically,kuhn2017full,kuhn2017optically,jin20216,rapp1967equatorial,demtroder2024astrophysics,hessels2006radio,shakura1973black,morgan2010quasar,bentz2015agn}}
    \label{tab:systems}
\end{table*}

The first two terms of the Hamiltonian in equation \eqref{eq:hamrel} comprise the standard Dirac Hamiltonian in flat spacetime, which we will denote by $H\lo{\text{Dirac}}$ in the following. The remaining terms arise from the non-inertial frame. More specifically, the third term corresponds to the spin-rotation coupling (a relativistic analog of the Mashhoon effect \cite{mashhoon1988neutron}), while the fourth and fifth term involve the centrifugal force, in terms of the centrifugal field coupling and a centrifugal redshift, respectively.

We note that our derivation is general and applies for relativistic (Dirac) and non-relativistic (Schrödinger) fermions. The Hamiltonian for the latter can be derived through two consecutive Foldy-Wouthuysen transformations~\cite{bjorken1965relativistic}, which we discuss further in the supplementary material. In the following, we focus on the effects of the inertial coupling to the centrifugal force and neglect other terms without loss of generality. This means we consider the relativistic Hamiltonian:
\begin{equation}
    H=H\lo{\mathrm{Dirac}}-\beta\gamma\hi{2}\bm{F}\lo{\text{centr}}\cdot\bm{r}.
    %H=\frac{\bm{p}\hi{2}}{2m}-\gamma\hi{2}\bm{F}\lo{centr}\cdot\bm{r},
\end{equation}
which describes a free Dirac particle subject to a uniform inertial field. In momentum space, the corresponding Lagrangian \cite{gergely2002hamiltonian} becomes;

\begin{equation}
    \mathcal{L}=\mathcal{L}\lo{\mathrm{Dirac}}+i\beta\gamma\hi{2}F\lo{\mathrm{centr}}\pkd{k}\nabla\lo{\bm{k}}\pk{k},
    %\mathcal{L}=\pkd{k}\left(\omega-\frac{k\hi{2}}{2m}\right)\pk{k}+i\gamma\hi{2}F\lo{\mathrm{centr}}\pkd{k}\nabla\lo{\bm{k}}\pk{k},
    \label{eq:lg1}
\end{equation}
where $\pk{k}$ denotes the fermion Dirac spinor at momentum $\bm{k}$. 

Requiring gauge invariance under a local momentum-dependent geometric $U(1)$ gauge transformation \mbox{$\pk{k}\mapsto e\hi{i\beta(\bm{k})}\pk{k}$}, we promote the momentum derivative to a covariant derivative \mbox{$\bm{D}\lo{\bm{k}}=\bm{\nabla}\lo{k}-ig\bm{\mathcal{A}}\lo{\bm{k}}$} by minimal coupling with the gauge potential $\bm{\mathcal{A}}\lo{\bm{k}}$. This prescription is analogous to the minimal coupling in quantum electrodynamics~\cite{griffiths2020introduction}. However, here, the connection is not electromagnetic but geometric in origin, emerging from the geometric structure of the non-inertial frame. In fact, the gauge potential $\bm{\mathcal{A}\lo{\bm{k}}}$ corresponds to the Berry potential~\cite{gangaraj2017berry,berry1984quantal} (see supplemental material for details). In further analogy to the quantum electromagnetic theory and from symmetry principles, we introduce an energy term for the free geometric field proportional to the square of the Berry curvature \mbox{$ \Omega\lo{\mn}=\epsilon\lo{\mn\sigma}(\bm{\nabla}\lo{k}\times\bm{\mathcal{A}}\lo{k})\lo{\sigma} $}, and a suitable gauge-fixing term to obtain the full model Lagrangian:

%\begin{widetext}
 %   \begin{equation}
 \begin{multline}
%    \mathcal{L}=(\omega+k\hi{2})\pkd{k}\pk{k}+i\gamma\hi{2}F\lo{\mathrm{centr}}\pkd{k}\left(\nabla\lo{\bm{k}}-ig\bm{\mathcal{A}}(\bm{k})\right)\pk{k}-\frac{1}{4}\Omega\lo{\mn}\Omega\hi{\mn}+\mathcal{L}\lo{gf}
    \mathcal{L}=\mathcal{L}\lo{\mathrm{Dirac}}+i\beta\gamma\hi{2}\pkd{k}\left[\mathbf{F}\lo{\mathrm{centr}}\cdot\left(\bm{\nabla}\lo{\bm{k}}-i\bm{\mathcal{A}}(\bm{k})\right)\right]\pk{k}\\-\frac{1}{4g\hi{2}}\Omega\lo{\mn}\Omega\hi{\mn}+\mathcal{L}\lo{\text{gf}},
    %\mathcal{L}=\left(\omega-\frac{k\hi{2}}{2m}\right)\pkd{k}\pk{k}+i\gamma\hi{2}\pkd{k}\left[\mathbf{F}\lo{\mathrm{centr}}\cdot\left(\nabla\lo{\bm{k}}-i\bm{\mathcal{A}}(\bm{k})\right)\right]\pk{k}\\-\frac{1}{4g\hi{2}}\Omega\lo{\mn}\Omega\hi{\mn}+\mathcal{L}\lo{gf},
    \label{eq:lgk}
\end{multline}

where we have introduced the geometric coupling constant $g$, which has units of $\unit{\newton}\hi{-\frac{1}{2}}$.
%\textcolor{red}{[SOMETHING DOESN'T WORK WITH THE UNITS HERE?]}
%\end{equation}
%\end{widetext}

The first term describes the Dirac particle, the second term encodes the coupling to the geometric gauge potential induced by the centrifugal force, and the final two terms describe the dynamics and gauge choice of the emergent gauge field. 

%\section{\label{sec:inertialrep}Inertial repulsion}
%

To derive the effective interaction term, called \textit{inertial repulsion}, we decompose the action corresponding to the Lagrangian in Eq. \eqref{eq:lgk} into \mbox{$S = S_0 + S_{\text{gauge}}$}, where $S_{\text{gauge}}$ denotes the geometric gauge-potential dependent part:
\begin{multline}
    S\lo{\text{gauge}}=\int\ud \omega \ud\hi{3}k \left\{\beta\gamma\hi{2}\bm{F}\lo{\text{centr}}\bm{\mathcal{A}}\lo{\bm{k}}\pkd{k}\pk{k}\right.\\-\left.\frac{1}{4g\hi{2}}\Omega\lo{\mn}\Omega\hi{\mn}+\mathcal{L}\lo{\text{gf}}\right\},
\end{multline}
and $S_0$ contains all other terms. % where a part depending on the geometric gauge potential  consider the gauge-dependent part of the action and integrate out the gauge field $\bm{\mathcal{A}}\lo{\bm{k}}$ in the path-integral formulation, where the exponential $\exp(iS\lo{gauge})$ of the action provides the weight of each field configuration. Since the terms in the action that do not explicitly depend on the gauge potential contribute only an overall multiplicative factor to the integral, it suffices to consider the explicitly gauge potential-dependent terms of the action:
Expanding the Berry curvature $\Omega\lo{\mn}$ in derivatives of the Berry potential and taking the gauge-fixing term to be Coulomb gauge \cite{magpantay1994coulomb}, i.e. $\mathcal{L}\lo{\text{gf}}=\left(\bm{\nabla}\lo{\bm{k}}\bm{\mathcal{A}}\lo{\bm{k}}\right)\hi{2}$, the gauge part of the action can be rewritten in quadratic form (see supplemental material for details):
\begin{equation}
    S\lo{\text{gauge}}=\int\ud \omega\ud k\hi{3}\left\{\beta\bm{J}\lo{\bm{k}}\bm{\mathcal{A}}\lo{\bm{k}}-\frac{1}{2}\bm{\mathcal{A}}\lo{\bm{k}}K\bm{\mathcal{A}}\lo{k}\right\}.
\end{equation}
Here, the effective current is given by $\bm{J}\lo{\bm{k}}=\gamma\hi{2}\bm{F}\lo{\mathrm{centr}}\pkd{k}\pk{k}$, and $K=-\eta g\hi{-2}\Box\lo{\bm{k}}$ is the kinetic operator for the gauge field, with the Minkowski metric $\eta=\mathrm{diag}(+1,-1,-1,-1)$ and the momentum-space d'Alembertian $\Box\lo{\bm{k}}$. 

The effective action is obtained by integrating out the gauge potential $\bm{\mathcal{A}}\lo{\bm{k}}$ over the exponential of the gauge-dependent action terms \cite{ferrero2025one,scharnhorst1997functional}, while the terms in the action that do not explicitly depend on the gauge potential, i.e., $S_0$, only contribute an overall multiplicative factor to the integral. This takes the form of a Gaussian integral~\cite{zee2010quantum}:
\begin{align}
   &e\hi{iS\lo{\text{gauge}}}\\&=\int\prod\limits\lo{\mu} \mathcal{D}\mathcal{A}\lo{\mu}\exp\left(i\int\ud \omega\ud\hi{3}k\left\{\beta\bm{J}\lo{\bm{k}}\bm{\mathcal{A}}\lo{\bm{k}}-\frac{1}{2}\bm{\mathcal{A}}\lo{\bm{k}}K\bm{\mathcal{A}}\lo{\bm{k}}\right\}\right)\nonumber\\&=\sqrt{\frac{(2\pi)\hi{4}}{\det(K)}}\exp\left(\frac{i}{2}\int\ud \omega\ud\hi{3}k\ud\hi{3}k'\bm{J}\lo{\bm{k}}D(\bm{k}-\bm{k'})\bm{J}\lo{\bm{k'}}\right),\nonumber
\end{align}
where the $D(\bm{q})=\eta\frac{g\hi{2}}{4\pi\vert\bm{q}\vert}$ is the Green's function (or propagator) defined by $KD(\bm{q})=\delta\hi{(3)}(\bm{q})$ and we have used that $\beta\hi{2}=\mathbf{1}$. 

Taking the logarithm yields the effective action, from which the effective Lagrangian can be determined. In momentum space, the emergent effective interaction term in the Lagrangian takes the form of:
\begin{equation}
    \mathcal{L}\lo{\text{int}}\hi{\text{geom}}=-U\lo{\text{rep}}\int\ud\hi{3}k' \rho(\bm{k})\frac{1}{4\pi\vert\bm{k}-\bm{k'}\vert}\rho(\bm{k'}),
    \label{eq:intk}
\end{equation}
with \mbox{$U\lo{\text{rep}}=\frac{1}{2}g\hi{2}\gamma\hi{4}F\lo{\text{centr}}\hi{2}$} and \mbox{$\rho(\bm{k})=\pkd{k}\pk{k}$}.

This interaction is non-local in momentum space and resembles a Coulomb-like repulsion. Here, $\rho(\bm{k})$ plays the role of a geometric charge density that sources the interaction. 

Fourier transformation back to position space yields the corresponding real-space interaction term, the inertial repulsion:
%\begin{widetext}
% \begin{equation}
%     \mathcal{L}\lo{int}\hi{geom}=-\frac{1}{2}(2\pi)\hi{3}\,g\hi{2}\gamma\hi{4}\,F\lo{\text{centr}}\hi{2}\,\int\ud\hi{3}r'\ud\hi{3}r''\psi\dg(\bm{r'})\psi(\bm{r'+r})\frac{1}{\vert\bm{r}\vert\hi{2}}\psi\dg(\bm{r''})\psi(\bm{r''-r}).
%     \label{eq:intr}
% \end{equation}
\begin{equation*}
    \mathcal{L}\lo{\text{int}}\hi{\text{geom}}=-(2\pi)\hi{3}U\lo{\text{rep}}\int\ud\hi{3}r'\ud\hi{3}r''\psi\dg\lo{\bm{r+r'}}\psi\lo{\bm{r'}}\frac{1}{\vert\bm{r}\vert\hi{2}}\psi\dg\lo{\bm{r''}}\psi\lo{\bm{r+r''}},
\end{equation*}
which can be rewritten as:
\begin{equation}
    \mathcal{L}\lo{\text{int}}\hi{\text{geom}}=-(2\pi)\hi{3}U\lo{\text{rep}}\frac{\vert\Gamma\lo{\bm{r}}\vert\hi{2}}{\vert \bm{r}\vert\hi{2}},
    \label{eq:intr}
\end{equation}
where $\Gamma\lo{\bm{r}}$ is given by the integral:
\begin{equation}
\Gamma\lo{\bm{r}}=\int\ud\hi{3} r' \psi\dg\lo{\bm{r}+\bm{r'}}\psi\lo{\bm{r'}}.
\end{equation}
%\end{widetext}
%\textcolor{red}{[I THINK THE UNIT ERROR FROM (4) PERSISTS UNTIL HERE. (10) IS NOT IN UNITS OF AN ENERGY DENSITY]}

This interaction is symmetric around the origin of rotation. The corresponding force can be obtained as $F=-\nabla U\propto \frac{1}{\vert\bm{r}\hi{3}\vert}$, which mirrors charge-dipole interaction in three dimensions~\cite{coon2002anomalies,griffiths2023introduction}. The interaction is repulsive, long-ranged, and emerges from the geometric structure of the rotating frame. 

Consequently, this interaction arises universally for fermions in rotating frames due to the coupling of the centrifugal field to the geometric Berry potential. 

Table \ref{tab:systems} shows the centrifugal force and interaction strength w.r.t. the coupling constant for electrons and neutrons in different rotating candidate systems. The strength of the interaction scales as \mbox{$g\hi{4}\gamma\hi{4}F\lo{\text{centr}}\hi{2}$}, where the centrifugal force $F\lo{\text{centr}}=m\omega\hi{2}d$ depends on the mass $m$ of the fermion, the angular velocity $\omega$, and the radius $d$ of the rotation. Thus, systems with large $\omega$ or $d$ will exhibit a significantly enhanced effective repulsive interaction. These include high-frequency, small-radius systems, such as rotating ions and molecules in, e.g., metal-organic frameworks, or astrophysical systems with large rotational radii, such as fermions in rapidly spinning neutron stars quasars and their accretion disks. 

As illustrated in Table \ref{tab:systems}, the effect becomes most significant for axial phonons, rotating small molecules, and rotating linker molecules in metal-organic frameworks. While the table captures the qualitative scaling of the inertial repulsion, a quantitative comparison of the effect for the different system is not possible as the value of the coupling constant $g$ cannot be obtained from the gauge theory alone and has to be fixed by comparing theoretical predictions of the inertial repulsion to measurable quantities. The interaction strengths in Table~\ref{tab:systems} should therefore be viewed as relative indicators of scaling rather than absolute values.

While the derivation of the inertial repulsion was motivated from rotation-induced inertial effects and is applicable to a wide range of rotating physical systems, the underlying mechanism extends far beyond purely rotational systems. We derived the emergent, repulsive, long-ranged interaction for a fermion in a rotating frame. This effective interaction arises after integrating out the geometric gauge potential (Berry potential) from the Lagrangian containing a centrifugal field coupling correction term due to the non-inertial frame of reference. The resulting effective interaction in Eq. \ref{eq:intk} shows a Coulomb-like structure in momentum space. In real space, this leads to a symmetric interaction term that decays as $1/\vert\bm{r}\vert\hi{2}$ and corresponds to a force $F\propto 1/\vert\bm{r}\vert\hi{3}$, mirroring a charge-dipole interaction. The necessary prerequisite for this interaction is the presence of an interaction term in the Hamiltonian that couples linearly to the position operator $\bm{r}$. Such terms can be Fourier transformed into momentum space, leading to derivatives in the momentum which can minimally couple to a geometric gauge field and, consequently, yield the repulsive effective interaction term derived in the previous section. This framework is therefore not restricted to pure rotation but can be applied to general accelerated systems with acceleration $\bm{a}$, where an inertial potential $\sim m\bm{a}\cdot\bm{r}$ linear in $\bm{r}$ appears. In fact, the derivation of the centrifugal coupling term in the Hamiltonian conducted by Hehl and Ni \cite{hehl1990inertial} does not rely only on rotation but instead derives the term for a general 4-acceleration of the observer. We expect to see the emergence of a similar inertial repulsion term, depending on the four-acceleration $a$ of the observer's frame. Further discussion of such a term lies beyond the scope of this paper and will be covered separately. The derivation presented in the supplementary material considers the special case $\bm{a}=-\omega\hi{2}\bm{d}$ of a purely rotational system with angular velocity $\omega$ and radial position $\bm{d}$. 

More generally, the formalism presented here can be extended to further fields coupling linearly to the position operator, leading to a repulsive effective geometric potential. A prominent example is the coupling of an electric field to a charged particle \cite{talin2008electron,griffiths2020introduction}, where the potential energy takes the role of the centrifugal field coupling, or infrared-active phonons \cite{decius1968dipolar,zan2024electron}, which involve electrons coupling to dipolar displacements. 

The inertial repulsion derived in this work, therefore, does not only constitute a peculiarity of the specific rotational system, but rather has wide-ranging potential applications as a geometric mechanism and can contribute to previously neglected corrections to fermionic energy levels in rotating frames.

\section*{Acknowledgements}
\begin{acknowledgments}
 We acknowledge support from the Knut and Alice Wallenberg Foundation (Grant No. 2023.0087), the Swedish Research Council (VR starting Grant No. 2022-03350), the Olle Engkvist Foundation (Grant No. 229-0443), the Royal Physiographic Society in Lund (Horisont), and Chalmers University of Technology, via the department of physics and the Areas of Advance Nano and Materials Science. 
\end{acknowledgments}


\clearpage
\newpage
\widetext

\begin{center}
\textbf{\large Supplementary Materials: Inertial Repulsion from Quantum Geometry}\\[1.2ex]
Maike Fahrensohn and R. Matthias Geilhufe\\[1.2ex]
\textit{Department of Physics, Chalmers University of Technology, 412 96 G\"{o}teborg, Sweden}
\end{center}

\section*{Inertial Effects in Rotating Ions
\label{app:inertial}}
Following Refs.~\cite{geilhufe2022dynamic} and~\cite{hehl1990inertial}, we consider a Dirac particle in a rotating system. This can, e.g., be an electron bound to a rotating ion~\cite{geilhufe2022dynamic} or a neutron in a rapidly-rotating neutron star. To describe the dynamics of the Dirac particle in the non-inertial, rotating frame of the observer, it is useful to introduce the comoving local orthonormal tetrads $e\lo{\alpha}$ that evolve with proper time $\tau$ according to the generalized Fermi-Walker transport law~\cite{hehl1991two,thorne2000gravitation,ni1977proper}: %the dynamics of an electron bound to a rotating ion can be described in the non-inertial rest frame of the electron co-moving with the ion. The local orthonormal tetrads $e\lo{\alpha}$ of this frame evolve with proper time $\tau$ according to the generalized Fermi-Walker transport law \cite{hehl1991two,thorne2000gravitation,ni1977proper} 
\begin{equation}
    \dv{e\lo{\alpha}}{\tau}=\Omega\lo{\mathrm{FW}}\cdot e\lo{\alpha}.
\end{equation}
Denoting the 4-acceleration of the observer by $a\hi{\mu}$, the 4-rotation $\omega\hi{\mu}$, and the 4-velocity with $u\hi{\mu}$, the generalized Fermi-Walker transport tensor takes the shape~\cite{geilhufe2022dynamic}:
\begin{equation}
    \Omega\lo{\mathrm{FW}}\hi{\mn}=\frac{1}{c\hi{2}}\left(a\hi{\mu}u\hi{\nu}-a\hi{\nu}u\hi{\mu}\right)+\vphantom{\frac{1}{c\hi{2}}}u\hi{\alpha}\omega\hi{\beta}\epsilon\hi{\alpha\beta\mn},
\end{equation}
and $e\lo{0}\hi{\mu}=u\hi{\mu}$, since the coordinate tetrad is chosen for the observer's rest frame. The generalized Fermi-Walker tensor is split into a non-rotating Fermi-Walker part and a rotational part. 

In the rotating frame, the Dirac equation takes the form~\cite{sciama1963recent,hehl1985kinematics,kibble1961lorentz,hehl1990inertial}:
\begin{equation}
    \gamma\hi{\alpha}i\hbar D\lo{\alpha}\Psi=mc \Psi,
\end{equation}
where $\Psi$ is the Dirac spinor, $\gamma\hi{\alpha}$ are the Dirac matrices, satisfying the anti-commutation relation \mbox{$\{\gamma\hi{\mu},\gamma\hi{\nu}\}=\eta\hi{\mn}$}, $m$ is the fermion mass, and the covariant derivative is given by:
\begin{equation}
    D\lo{\alpha}=\partial\lo{\alpha}-\frac{i}{4}\sigma\hi{bc}\Gamma\lo{bca}.
\end{equation}

Here, $\sigma\hi{\beta\delta}=\frac{i}{2}[\gamma\hi{\beta},\gamma\hi{\delta}]$, and $\Gamma\lo{bca}$ are the connection coefficients, constructed from the moments of anholonomicity of the tetrads. 

The covariant Dirac equation can then be recast into Schrödinger form:
\begin{equation}
    i\hbar\dv{\Psi}{t}=H\Psi,
\end{equation}
with the Hamiltonian~\cite{geilhufe2022dynamic}:
\begin{align}
    H&=\beta mc\hi{2}+c\bm{\alpha}\cdot\bm{p}-\bm{\omega}\cdot \bm{J}\\
    &-\beta\gamma\hi{2}\bm{F}\lo{\text{centr}}\cdot\bm{r}-\frac{\gamma\hi{2}}{2mc}\{\bm{F}\lo{\text{centr}}\cdot\bm{r},\bm{p}\cdot\bm{\alpha}\},
\end{align}
where $\beta$ and $\bm{\alpha}$ are Dirac matrices, $\bm{p}$ and $\bm{r}$ denote the momentum and position operators, $\bm{\omega}$ is the angular velocity of the rotating frame, $\bm{J}=\bm{L}+\bm{S}$ is the total angular momentum, and the centrifugal force is given by $\bm{F}\lo{\text{centr}}=m\omega\hi{2}\bm{d}$.

The first two terms correspond to the conventional Dirac Hamiltonian in flat spacetime. The additional inertial terms arising from rotation are described in Table \ref{tab:relinertialterms}.

\begin{table}[b]
    \centering
    \begin{tabular}{cc}
    \hline\hline
      \text{Spin-rotation coupling} &$\bm{\omega}\cdot\bm{J}$\\
       \text{Centrifugal field coupling}  & $\beta\gamma\hi{2}\bm{F}\lo{\text{centr}}\cdot\bm{r}$\\
       \text{energy-momentum redshift}& $\frac{\gamma\hi{2}}{2mc}\{\bm{F}\lo{\text{centr}}\cdot\bm{r},\bm{p}\cdot\bm{\alpha}\}$ \\\hline\hline
    \end{tabular}
    \caption{Relativistic inertial effects in the Hamiltonian for a Dirac particle in a rotating frame\cite{hehl1990inertial}}
    \label{tab:relinertialterms}
\end{table}

\begin{table}[t]
    \centering
    \begin{tabular}{cc}
    \hline\hline
      \text{Spin-rotation coupling}   & $\bm{\omega}\cdot\bm{J}$\\
       \text{Centrifugal field coupling}  & $\gamma\hi{2}\bm{F}\lo{\text{centr}}\cdot\bm{r}$\\
       \text{Centrifugal spin-orbit coupling} & $\frac{\gamma\hi{2}}{2m\hi{2}c\hi{2}}\bm{F}\lo{\text{centr}}\cdot(\bm{S}\times\bm{p})$\\
       \text{Centrifugal redshift} & $\frac{\gamma\hi{2}}{2m\hi{2}c\hi{2}}\bm{p}(\bm{F}\lo{\text{centr}}\cdot\bm{r})\bm{p}$\\\hline\hline
    \end{tabular}
    \caption{Inertial terms arising in the non-relativistic Hamiltonian for a fermion in a rotating frame \cite{geilhufe2022dynamic}}
    \label{tab:nonrelinertialterms}
\end{table}

Through two consectutive Foldy-Wouthuysen transformations \cite{bjorken1965relativistic}, evaluated up to $(mc\hi{2})\hi{-1}$, and removal of the rest mass, the Hamiltonian can be brought into a non-relativistic form \cite{geilhufe2022dynamic}:
\begin{align}
    H&=\frac{\bm{p}\hi{2}}{2m}-\bm{\omega}\cdot\bm{J}-\gamma\hi{2}\bm{F}\lo{\text{centr}}\cdot\bm{r}\nonumber\\
    &-\frac{\gamma\hi{2}}{2m\hi{2}c\hi{2}}\left(\bm{p}(\bm{F}\lo{\text{centr}}\cdot\bm{r})\bm{p}+\bm{F}\lo{\text{centr}}\cdot(\bm{S}\times\bm{p}\right),
    \label{eq:inertialham}
\end{align}
with the Lorentz factor $\gamma=\left(1-\frac{d\hi{2}\omega\hi{2}}{c\hi{2}}\right)\hi{-\frac{1}{2}}\approx1$.

A summary of the inertial terms and their physical interpretation is provided in Table \ref{tab:nonrelinertialterms}.

In the derivation of the inertial repulsion, we focus exclusively on the contributions of the centrifugal field coupling term and work in natural units. 

\section*{\label{app:lagrangian}Some Remarks on the Model Lagrangian}
We consider a relativistic Hamiltonian of the form
\begin{equation}
    H=H\lo{\text{Dirac}}-\beta\gamma\hi{2}\bm{F}\lo{\text{centr}}\cdot\bm{r},
\end{equation}
consisting of a standard Dirac Hamiltonian $H\lo{\text{Dirac}}$ and a term corresponding to the centrifugal field coupling in Eq. \ref{eq:inertialham}.

The Lagrangian in momentum space can then be obtained as:
\begin{equation}
    \mathcal{L}=\mathcal{L}\lo{\text{Dirac}}+i\beta\gamma\hi{2}\bm{F}\lo{\mathrm{centr}}\pkd{k}\bm{\nabla}\lo{\bm{k}}\pk{k}.
    %\mathcal{L}=\omega\pkd{k}\pk{k}-\frac{k\hi{2}}{2m}\pkd{k}\pk{k}+i\gamma\hi{2}\bm{F}\lo{\mathrm{centr}}\pkd{k}\bm{\nabla}\lo{\bm{k}}\pk{k},
    \label{eq:lginertial}
\end{equation} 

Let us now consider a momentum-dependent geometric $U(1)$ gauge transformation
\begin{equation}
    \pk{k}\mapsto e\hi{i\beta(\bm{k})}\pk{k}, 
\end{equation}
arising from the local geometric structure of the Hilbert space. 

The corresponding gauge potential $\bm{\mathcal{A}}\lo{\bm{k}}$ then transforms according to
\begin{equation}
    \bm{\mathcal{A}}\lo{\bm{k}}\mapsto \bm{\mathcal{A}}\lo{\bm{k}} +\bm{\nabla}\lo{\bm{k}}\beta(\bm{k}).
    \label{eq:geompot}
\end{equation}
In fact, this is fulfilled by the Berry connection \cite{gangaraj2017berry} $\bm{\mathcal{A}}\lo{\bm{k}}$ in terms of the Dirac spinor and the invariant field strength tensor is then given by the Berry curvature \cite{gangaraj2017berry}:
\begin{equation}
    \Omega\lo{\mn}=\partial\lo{\mu}\mathcal{A}\lo{\nu}-\partial\lo{\nu}\mathcal{A}\lo{\mu}=\epsilon\lo{\mn\sigma}\left(\bm{\nabla}\lo{\bm{k}}\times\bm{\mathcal{A}}\lo{\bm{k}}\right)\lo{\sigma},
\end{equation}
where $\partial\lo{\mu}$ denotes the partial derivative with respect to $k\lo{\mu}$.

To promote this to a full gauge theory, we replace the derivative in Eq. \ref{eq:lginertial} with a covariant derivative in momentum space, and include a kinetic term for the gauge field and a gauge-fixing term in the Lagrangian, resulting in:
%\begin{widetext}
    \begin{equation}
    \mathcal{L}=\mathcal{L}\lo{\text{Dirac}}+i\beta\gamma\hi{2}\bm{F}\lo{\mathrm{centr}}\pkd{k}\left(\bm{\nabla}\lo{\bm{k}}-i\bm{\mathcal{A}}\lo{\bm{k}}\right)\pk{k}-\frac{1}{4g\hi{2}}\Omega\lo{\mn}\Omega\hi{\mn}+\mathcal{L}\lo{\text{gf}}.
    %\left(\omega-\frac{k\hi{2}}{m}\right)\pkd{k}\pk{k}+i\gamma\hi{2}\bm{F}\lo{\mathrm{centr}}\pkd{k}\left(\bm{\nabla}\lo{\bm{k}}-i\bm{\mathcal{A}}\lo{\bm{k}}\right)\pk{k}-\frac{1}{4g\hi{2}}\Omega\lo{\mn}\Omega\hi{\mn}+\mathcal{L}\lo{gf}.
    \label{eq:lgk2}
\end{equation}
%\end{widetext}
\section*{\label{app:geomterms}Motivating the geometric terms}

% current motivation for Berry connection
Here, we provide additional context motivating the inclusion of the Berry connection as a geometrical gauge field and the emergence of a dynamic term for this gauge field in terms of the Berry curvature. 

Following Xiao et al.~\cite{xiao2009polarization}, we consider the first-order correction to the polarization current for an adiabatically varying centrifugal force $F=F\lo{\text{centr}}$. For a single band, this correction can be expressed as:
\begin{align}
    \int\ud tj\lo{\alpha}\hi{(1)}&=i\int\ud t\Dot{F}\lo{\beta}\int\frac{\ud\hi{3}k}{(2\pi)\hi{3}}\left(\pdv{}{k\lo{\alpha}}\langle\psi\vert\pdv{}{F\lo{\beta}}\vert\psi\rangle-\pdv{}{F\lo{\beta}}\langle\psi\vert\pdv{}{k\lo{\alpha}}\vert\psi\rangle\right)\\
    &=-i\int \ud t\int\frac{\ud\hi{3}k}{(2\pi)\hi{3}}\left(\pdv{}{t}\underbrace{\langle\psi\vert\pdv{}{k\lo{\alpha}}\vert\psi\rangle}\lo{=-i\mathcal{A}\lo{\alpha}}\right)\\
    &=-\int\ud t\int\frac{\ud\hi{3}k}{(2\pi)\hi{3}}\mathcal{A}\lo{\alpha},
\end{align}
where $\mathbf{\mathcal{A}}\lo{\mathbf{k}}$ denotes the Berry connection in momentum space. 

Including this correction modifies the position operator to:
\begin{equation}
    \mathbf{r}=
    %-\int\ud t \mathbf{j}\hi{(0)}-\int\ud t\mathbf{j}\hi{(1)}
    i\mathbf{\nabla}\lo{\mathbf{k}}+\mathbf{\mathcal{A}}\lo{\mathbf{k}},
\end{equation}
and motivates the promotion of the ordinary momentum space derivative to a covariant derivative $D\lo{\alpha}=\partial\lo{\alpha}-i\mathcal{A}\lo{\alpha}$. Here, $\partial\lo{\alpha}$ denotes the derivative with respect to $k\lo{\alpha}$ and $\mathcal{A}\lo{\alpha}$ is the $\alpha$-component of the Berry connection $\mathbf{\mathcal{A}}\lo{\mathbf{k}}$. This makes the gauge character of the Berry connection manifest. For a more detailed discussion on the gauge field properties of the Berry connection, we refer to Xiao et al.~\cite{xiao2010berry}.

The Berry curvature $\Omega\lo{\mathbf{k}}=\mathbf{\nabla}\lo{\mathbf{k}}\times\mathbf{\mathcal{A}}\lo{\mathbf{k}}$ acts as an effective magnetic field in momentum space. In analogy with classical Landau theory of magnetization~\cite{landau1937theory}, it is then natural to consider the inclusion of an energetic contribution
\begin{equation}
    \mathcal{L}\lo{\text{magn}}=\frac{\mu\lo{0}}{\chi} \bm{M}\hi{2}
\end{equation}
in the Lagrangian.

According to the modern theory of orbital magnetization \cite{thonhauser2011theory,resta2010electrical}, the local magnetization $M$ receives a correction stemming from the Berry curvature $\bm{\Omega}(\bm{q})$ that takes the form:

\begin{equation}
    \bm{M}(r)=\frac{e}{2\hbar}\int\limits\lo{BZ}\frac{\ud\hi{3}q}{(2\pi)\hi{3}}\omega(\bm{q},\bm{r})\bm{\Omega}(\bm{q}),
\end{equation}
where $\omega(\bm{q},\bm{r})$ is a weight that can depend on the Fermi energy. 

In momentum space, this becomes:
\begin{equation}
    \bm{M}(\bm{k})=\frac{e}{2\hbar}\int\limits\lo{BZ}\frac{\ud\hi{3}q}{(2\pi)\hi{3}}\bm{\Omega}(\bm{q})\underbrace{\int\ud\hi{3}r e\hi{-i\bm{kr}}\omega(\bm{q},\bm{r})}_{\eqqcolon \omega(\bm{q},\bm{k})}.
\end{equation}

The square of the magnetization term in the Lagrangian can then be written as:
\begin{align}
    \frac{\mu\lo{0}}{\chi}\int\ud\hi{3}r\bm{M}(\bm{r})\hi{2}&=\frac{\mu\lo{0}}{\chi}\int\frac{\ud\hi{3}k}{(2\pi)\hi{3}}\bm{M}(\bm{k})\bm{M}(-\bm{k})\\
    &=\frac{\mu\lo{0}}{\chi} \left(\frac{e}{2\hbar}\right)\hi{2}\int\frac{\ud\hi{3}q\ud\hi{3}q'}{(2\pi)\hi{6}}\omega(\bm{q},\bm{k})\omega(\bm{q'},-\bm{k})\bm{\Omega}(\bm{q})\bm{\Omega}(\bm{q'}).
\end{align}

In the long-wavelength limit, $\omega(\bm{q},\bm{r})$ varies slowly in $\bm{r}$ such that:
%$\omega(q,k)$ is sharply peaked around $k$ and can be approximated as 
\begin{equation}
    \omega(\bm{q},\bm{k})\approx(2\pi)\hi{3}\delta\hi{(3)}(k)\omega(\bm{q}),\quad \omega(\bm{q})=\omega(\bm{q},\bm{k}=\bm{0}).
\end{equation}

This approximation yields:
\begin{equation}
    \frac{\mu\lo{0}}{\chi}\int\ud\hi{3}r\bm{M}(\bm{r})\hi{2}=\frac{\mu\lo{0}}{\chi}\left(\frac{e}{2\hbar}\right)\hi{2}\int\frac{\ud\hi{3}q\ud\hi{3}q'}{(2\pi)\hi{3}}\omega(\bm{q})\omega(\bm{q'})\bm{\Omega}(\bm{q})\bm{\Omega}(\bm{q'}).
\end{equation}
Assuming diagonal kernel approximation, this reduces to:
\begin{equation}
    \frac{\mu\lo{0}}{\chi}\int\ud\hi{3}r\bm{M}(\bm{r})\hi{2}=\frac{\mu\lo{0}}{\chi}\left(\frac{e}{2\hbar}\right)\hi{2}\int\frac{\ud\hi{3}q}{(2\pi)\hi{3}} \omega\hi{2}(\bm{q})\Omega\lo{\mn}\Omega\hi{\mn}.
\end{equation}
The Lagrangian then acquires the additional geometric term:
\begin{equation}
    \mathcal{L}\supset -\frac{\mu\lo{0}}{\chi}\left(\frac{e}{2\hbar}\right)\hi{2}\omega\hi{2}(\bm{q})\Omega\lo{\mn}\Omega\hi{\mn}.
\end{equation}
To lowest order, $\omega$ may be regarded as constant, leading to a Lagrangian distribution mirroring the form of the field-strength tensor term $-\frac{1}{4g\hi{2}}F\lo{\mn}F\hi{\mn}$ in gauge theory. This reinforces the interpretation of the Berry curvature as a geometric field strength with the Berry connection as the corresponding gauge field.

\section*{\label{app:effaction}Effective Action Calculation}
We begin with the full Lagrangian in Eq. \ref{eq:lgk2}. To isolate the effects of the geometric gauge field, we consider its contributions to the action. It suffices to study the explicitly gauge potential-dependent terms in the action:
\begin{equation}
    S\lo{\text{gauge}}=\int\frac{\ud \omega \ud\hi{3}k}{(2\pi)\hi{4}} \left\{\beta\gamma\hi{2}\bm{F}\lo{\mathrm{centr}}\bm{\mathcal{A}}\lo{\bm{k}}\pkd{k}\pk{k}-\frac{1}{4g\hi{2}}\Omega\lo{\mn}\Omega\hi{\mn}+\mathcal{L}\lo{\text{gf}}\right\}.
\end{equation}

To obtain the quadratic structure of this gauge sector, we first rewrite the square of the Berry curvature through integration by parts:
\begin{align}
    &\frac{1}{2}\int\frac{\ud\omega\ud\hi{3}k}{(2\pi)\hi{4}}\Omega\lo{\mn}\Omega\hi{\mn}\\
    &=\int\frac{\ud\omega\ud\hi{3}k}{(2\pi)\hi{4}}\left(\dm{\mu}\mathcal{A}
    %\ba{}
    \lo{\nu}\partial\hi{\mu}\mathcal{A}\hi{\nu}-\dm{\mu}\mathcal{A}\lo{\nu}\partial\hi{\nu}\mathcal{A}\hi{\mu}
    \right)\nonumber\\
    %&=-\int\frac{\ud\omega\ud\hi{3}k}{(2\pi)\hi{4}}\left(\bm{\mathcal{A}}\lo{\nu}\dm{\mu}\partial\hi{\mu}\bm{\mathcal{A}}\hi{\nu}-\bm{\mathcal{A}}\lo{\nu}\dm{\mu}\partial\hi{\nu}\bm{\mathcal{A}}\hi{\mu}\right)\\
    &=-\int\frac{\ud\omega\ud\hi{3}k}{(2\pi)\hi{4}}\mathcal{A}\lo{\mu}\left(\dm{\nu}\partial\hi{\nu}\mathcal{A}\hi{\mu}-\dm{\nu}\partial\hi{\mu}\mathcal{A}\hi{\nu}\right)\nonumber\\
    &=-\int\frac{\ud\omega\ud\hi{3}k}{(2\pi)\hi{3}}\mathcal{A}\lo{\mu}\left(\Box\lo{\bm{k}}\eta\hi{\mn}-\partial\hi{\mu}\partial\hi{\nu}\right)\mathcal{A}\lo{\nu}\nonumber.
\end{align}
The last term can be cancelled by using Coulomb gauge~\cite{magpantay1994coulomb} as the gauge fixing term in the Lagrangian, i.e. $\mathcal{L}\lo{\text{gf}}=g\hi{-2}\left(\bm{\nabla}\lo{k}\bm{\mathcal{A}}\lo{\bm{k}}\right)\hi{2} $. The gauge-dependent part of the action then simplifies to:
\begin{equation}
    S\lo{\text{gauge}}=\int\frac{\ud \omega\ud k\hi{3}}{(2\pi)\hi{4}}\left\{\beta\bm{J}\lo{\bm{k}}\bm{\mathcal{A}}\lo{\bm{k}}-\frac{1}{2}\bm{\mathcal{A}}\lo{\bm{k}}K\bm{\mathcal{A}}\lo{k}\right\},
\end{equation}
where we have introduced the effective current operator $\bm{J}\lo{\bm{k}}=\gamma\hi{2}\bm{F}\lo{\mathrm{centr}}\pkd{k}\pk{k}$ and the kinetic operator $K=-g\hi{-2}\eta\Box\lo{\bm{k}}$ of the gauge field, where $\eta=\mathrm{diag}(+1,-1,-1,-1)$ is the Minkowski metric tensor and $\Box\lo{\bm{k}}$ the momentum-space d'Alembertian. 

We now formally integrate out the gauge field over the exponential of the gauge-dependent action. 
Since the gauge sector is quadratic, this is a Gaussian integral \cite{zee2010quantum}:
\begin{align}
   e\hi{iS\lo{\text{gauge}}}&=\int \mathcal{D}\mathcal{A}\exp\left(i\int\frac{\ud \omega\ud\hi{3}k}{(2\pi)\hi{4}}\left\{\beta\bm{J}\lo{\bm{k}}\bm{\mathcal{A}}\lo{\bm{k}}-\frac{1}{2}\bm{\mathcal{A}}\lo{\bm{k}}K\bm{\mathcal{A}}\lo{\bm{k}}\right\}\right)\\ &=\sqrt{\frac{(2\pi)\hi{4}}{\det(K)}}\exp\left(\frac{i}{2}\int\frac{\ud \omega\ud\omega'\ud\hi{3}k\ud\hi{3}k'}{(2\pi)\hi{4}}\bm{J}\lo{\bm{k}}D(\bm{k}-\bm{k'})\bm{J}\lo{\bm{k'}}\right)\nonumber,
\end{align}
where we use $\beta\hi{2}=\mathbf{1}$ and introduce the propagator $D(\bm{k})$, defined as the inverse of $K$, i.e., $KD(\bm{k})=\delta\hi{(3)}(\bm{k}) $. This can be calculated to be:

\begin{align}
    KD(k)&=-\eta g\hi{-2}\Box\lo{\bm{k}}\int\ud\hi{3}re\hi{-i\bm{k}\bm{r}}\Tilde{D}(\bm{r})\\
    &=\eta g\hi{-2}\int\ud\hi{3}r r\hi{2}\Tilde{D}(\bm{r})e\hi{-i\bm{k}\bm{r}}\nonumber\\
    &\overset{!}{=}\delta\hi{(3)}(\bm{k})=\int\frac{\ud\hi{3}r}{(2\pi)\hi{3}}e\hi{-i\bm{k}\bm{r}}\nonumber,
\end{align}
and, therefore,
\begin{equation}
    D(\bm{k})=\eta g\hi{2}\int \frac{\ud\hi{3}r}{(2\pi)\hi{3}}\frac{e\hi{-i\bm{k}\bm{r}}}{r\hi{2}}=\eta\frac{g\hi{2}}{4\pi k}.
\end{equation}
Fourier transformation of the propagator to position space then yields:
\begin{equation}
    D(\bm{r})=\eta\frac{g\hi{2}}{(2\pi)\hi{3}r\hi{2}},
\end{equation}
which decays with $\frac{1}{r\hi{2}}$.


\begin{thebibliography}{68}%
\makeatletter
\providecommand \@ifxundefined [1]{%
 \@ifx{#1\undefined}
}%
\providecommand \@ifnum [1]{%
 \ifnum #1\expandafter \@firstoftwo
 \else \expandafter \@secondoftwo
 \fi
}%
\providecommand \@ifx [1]{%
 \ifx #1\expandafter \@firstoftwo
 \else \expandafter \@secondoftwo
 \fi
}%
\providecommand \natexlab [1]{#1}%
\providecommand \enquote  [1]{``#1''}%
\providecommand \bibnamefont  [1]{#1}%
\providecommand \bibfnamefont [1]{#1}%
\providecommand \citenamefont [1]{#1}%
\providecommand \href@noop [0]{\@secondoftwo}%
\providecommand \href [0]{\begingroup \@sanitize@url \@href}%
\providecommand \@href[1]{\@@startlink{#1}\@@href}%
\providecommand \@@href[1]{\endgroup#1\@@endlink}%
\providecommand \@sanitize@url [0]{\catcode `\\12\catcode `\$12\catcode `\&12\catcode `\#12\catcode `\^12\catcode `\_12\catcode `\%12\relax}%
\providecommand \@@startlink[1]{}%
\providecommand \@@endlink[0]{}%
\providecommand \url  [0]{\begingroup\@sanitize@url \@url }%
\providecommand \@url [1]{\endgroup\@href {#1}{\urlprefix }}%
\providecommand \urlprefix  [0]{URL }%
\providecommand \Eprint [0]{\href }%
\providecommand \doibase [0]{https://doi.org/}%
\providecommand \selectlanguage [0]{\@gobble}%
\providecommand \bibinfo  [0]{\@secondoftwo}%
\providecommand \bibfield  [0]{\@secondoftwo}%
\providecommand \translation [1]{[#1]}%
\providecommand \BibitemOpen [0]{}%
\providecommand \bibitemStop [0]{}%
\providecommand \bibitemNoStop [0]{.\EOS\space}%
\providecommand \EOS [0]{\spacefactor3000\relax}%
\providecommand \BibitemShut  [1]{\csname bibitem#1\endcsname}%
\let\auto@bib@innerbib\@empty
%</preamble>
\bibitem [{\citenamefont {Demtr{\"o}der}(2017)}]{demtroder2017mechanics}%
  \BibitemOpen
  \bibfield  {author} {\bibinfo {author} {\bibfnamefont {W.}~\bibnamefont {Demtr{\"o}der}},\ }\href {https://doi.org/10.1007/978-3-319-27877-3} {\emph {\bibinfo {title} {Mechanics and thermodynamics}}}\ (\bibinfo  {publisher} {Springer},\ \bibinfo {year} {2017})\BibitemShut {NoStop}%
\bibitem [{\citenamefont {Corben}\ and\ \citenamefont {Stehle}(2013)}]{corben2013classical}%
  \BibitemOpen
  \bibfield  {author} {\bibinfo {author} {\bibfnamefont {H.~C.}\ \bibnamefont {Corben}}\ and\ \bibinfo {author} {\bibfnamefont {P.}~\bibnamefont {Stehle}},\ }\href@noop {} {\emph {\bibinfo {title} {Classical mechanics}}}\ (\bibinfo  {publisher} {Courier Corporation},\ \bibinfo {year} {2013})\BibitemShut {NoStop}%
\bibitem [{\citenamefont {Hehl}\ and\ \citenamefont {Ni}(1990)}]{hehl1990inertial}%
  \BibitemOpen
  \bibfield  {author} {\bibinfo {author} {\bibfnamefont {F.~W.}\ \bibnamefont {Hehl}}\ and\ \bibinfo {author} {\bibfnamefont {W.-T.}\ \bibnamefont {Ni}},\ }\bibfield  {title} {\bibinfo {title} {Inertial effects of a dirac particle},\ }\href {https://doi.org/10.1103/PhysRevD.42.2045} {\bibfield  {journal} {\bibinfo  {journal} {Physical Review D}\ }\textbf {\bibinfo {volume} {42}},\ \bibinfo {pages} {2045} (\bibinfo {year} {1990})}\BibitemShut {NoStop}%
\bibitem [{\citenamefont {Matsuo}\ \emph {et~al.}(2011{\natexlab{a}})\citenamefont {Matsuo}, \citenamefont {Ieda}, \citenamefont {Saitoh},\ and\ \citenamefont {Maekawa}}]{Matsuo2011}%
  \BibitemOpen
  \bibfield  {author} {\bibinfo {author} {\bibfnamefont {M.}~\bibnamefont {Matsuo}}, \bibinfo {author} {\bibfnamefont {J.}~\bibnamefont {Ieda}}, \bibinfo {author} {\bibfnamefont {E.}~\bibnamefont {Saitoh}},\ and\ \bibinfo {author} {\bibfnamefont {S.}~\bibnamefont {Maekawa}},\ }\bibfield  {title} {\bibinfo {title} {Effects of mechanical rotation on spin currents},\ }\href {https://doi.org/10.1103/PhysRevLett.106.076601} {\bibfield  {journal} {\bibinfo  {journal} {Physical Review Letters}\ }\textbf {\bibinfo {volume} {106}},\ \bibinfo {pages} {076601} (\bibinfo {year} {2011}{\natexlab{a}})}\BibitemShut {NoStop}%
\bibitem [{\citenamefont {Matsuo}\ \emph {et~al.}(2011{\natexlab{b}})\citenamefont {Matsuo}, \citenamefont {Ieda}, \citenamefont {Saitoh},\ and\ \citenamefont {Maekawa}}]{Matsuo2011prb}%
  \BibitemOpen
  \bibfield  {author} {\bibinfo {author} {\bibfnamefont {M.}~\bibnamefont {Matsuo}}, \bibinfo {author} {\bibfnamefont {J.}~\bibnamefont {Ieda}}, \bibinfo {author} {\bibfnamefont {E.}~\bibnamefont {Saitoh}},\ and\ \bibinfo {author} {\bibfnamefont {S.}~\bibnamefont {Maekawa}},\ }\bibfield  {title} {\bibinfo {title} {Spin-dependent inertial force and spin current in accelerating systems},\ }\href {https://doi.org/10.1103/PhysRevB.84.104410} {\bibfield  {journal} {\bibinfo  {journal} {Physical Review B}\ }\textbf {\bibinfo {volume} {84}},\ \bibinfo {pages} {104410} (\bibinfo {year} {2011}{\natexlab{b}})}\BibitemShut {NoStop}%
\bibitem [{\citenamefont {Geilhufe}(2022)}]{geilhufe2022dynamic}%
  \BibitemOpen
  \bibfield  {author} {\bibinfo {author} {\bibfnamefont {R.~M.}\ \bibnamefont {Geilhufe}},\ }\bibfield  {title} {\bibinfo {title} {Dynamic electron-phonon and spin-phonon interactions due to inertia},\ }\href {https://doi.org/10.1103/PhysRevResearch.4.L012004} {\bibfield  {journal} {\bibinfo  {journal} {Physical Review Research}\ }\textbf {\bibinfo {volume} {4}},\ \bibinfo {pages} {L012004} (\bibinfo {year} {2022})}\BibitemShut {NoStop}%
\bibitem [{\citenamefont {Thorne}\ \emph {et~al.}(2000)\citenamefont {Thorne}, \citenamefont {Misner},\ and\ \citenamefont {Wheeler}}]{thorne2000gravitation}%
  \BibitemOpen
  \bibfield  {author} {\bibinfo {author} {\bibfnamefont {K.~S.}\ \bibnamefont {Thorne}}, \bibinfo {author} {\bibfnamefont {C.~W.}\ \bibnamefont {Misner}},\ and\ \bibinfo {author} {\bibfnamefont {J.~A.}\ \bibnamefont {Wheeler}},\ }\href {https://doi.org/10.1002/piuz.19750060209} {\emph {\bibinfo {title} {Gravitation}}}\ (\bibinfo  {publisher} {Freeman San Francisco},\ \bibinfo {year} {2000})\BibitemShut {NoStop}%
\bibitem [{\citenamefont {Ni}\ and\ \citenamefont {Zimmermann}(1978)}]{ni1978inertial}%
  \BibitemOpen
  \bibfield  {author} {\bibinfo {author} {\bibfnamefont {W.-T.}\ \bibnamefont {Ni}}\ and\ \bibinfo {author} {\bibfnamefont {M.}~\bibnamefont {Zimmermann}},\ }\bibfield  {title} {\bibinfo {title} {Inertial and gravitational effects in the proper reference frame of an accelerated, rotating observer},\ }\href {https://doi.org/10.1103/PhysRevD.17.1473} {\bibfield  {journal} {\bibinfo  {journal} {Physical Review D}\ }\textbf {\bibinfo {volume} {17}},\ \bibinfo {pages} {1473} (\bibinfo {year} {1978})}\BibitemShut {NoStop}%
\bibitem [{\citenamefont {DeFacio}\ \emph {et~al.}(1978)\citenamefont {DeFacio}, \citenamefont {Dennis},\ and\ \citenamefont {Retzloff}}]{defacio1978presymmetry}%
  \BibitemOpen
  \bibfield  {author} {\bibinfo {author} {\bibfnamefont {B.}~\bibnamefont {DeFacio}}, \bibinfo {author} {\bibfnamefont {P.~W.}\ \bibnamefont {Dennis}},\ and\ \bibinfo {author} {\bibfnamefont {D.~G.}\ \bibnamefont {Retzloff}},\ }\bibfield  {title} {\bibinfo {title} {Presymmetry of classical relativistic particles},\ }\href {https://doi.org/https://journals.aps.org/prd/abstract/10.1103/PhysRevD.18.2813} {\bibfield  {journal} {\bibinfo  {journal} {Physical Review D}\ }\textbf {\bibinfo {volume} {18}},\ \bibinfo {pages} {2813} (\bibinfo {year} {1978})}\BibitemShut {NoStop}%
\bibitem [{\citenamefont {Page}(1975)}]{Page1975}%
  \BibitemOpen
  \bibfield  {author} {\bibinfo {author} {\bibfnamefont {L.~A.}\ \bibnamefont {Page}},\ }\bibfield  {title} {\bibinfo {title} {Effect of earth's rotation in neutron interferometry},\ }\href {https://doi.org/10.1103/PhysRevLett.35.543} {\bibfield  {journal} {\bibinfo  {journal} {Physical Review Letters}\ }\textbf {\bibinfo {volume} {35}},\ \bibinfo {pages} {543} (\bibinfo {year} {1975})}\BibitemShut {NoStop}%
\bibitem [{\citenamefont {Mashhoon}(1988)}]{mashhoon1988neutron}%
  \BibitemOpen
  \bibfield  {author} {\bibinfo {author} {\bibfnamefont {B.}~\bibnamefont {Mashhoon}},\ }\bibfield  {title} {\bibinfo {title} {Neutron interferometry in a rotating frame of reference},\ }\href {https://doi.org/10.1103/PhysRevLett.61.2639} {\bibfield  {journal} {\bibinfo  {journal} {Physical review letters}\ }\textbf {\bibinfo {volume} {61}},\ \bibinfo {pages} {2639} (\bibinfo {year} {1988})}\BibitemShut {NoStop}%
\bibitem [{\citenamefont {Werner}\ \emph {et~al.}(1979)\citenamefont {Werner}, \citenamefont {Staudenmann},\ and\ \citenamefont {Colella}}]{Werner1979}%
  \BibitemOpen
  \bibfield  {author} {\bibinfo {author} {\bibfnamefont {S.~A.}\ \bibnamefont {Werner}}, \bibinfo {author} {\bibfnamefont {J.~L.}\ \bibnamefont {Staudenmann}},\ and\ \bibinfo {author} {\bibfnamefont {R.}~\bibnamefont {Colella}},\ }\bibfield  {title} {\bibinfo {title} {Effect of earth's rotation on the quantum mechanical phase of the neutron},\ }\href {https://doi.org/10.1103/PhysRevLett.42.1103} {\bibfield  {journal} {\bibinfo  {journal} {Physical Review Letters}\ }\textbf {\bibinfo {volume} {42}},\ \bibinfo {pages} {1103} (\bibinfo {year} {1979})}\BibitemShut {NoStop}%
\bibitem [{\citenamefont {Huang}\ and\ \citenamefont {Geilhufe}(2024)}]{Huang2024}%
  \BibitemOpen
  \bibfield  {author} {\bibinfo {author} {\bibfnamefont {Z.}~\bibnamefont {Huang}}\ and\ \bibinfo {author} {\bibfnamefont {R.~M.}\ \bibnamefont {Geilhufe}},\ }\bibfield  {title} {\bibinfo {title} {Quantum metal‐organic frameworks},\ }\bibfield  {journal} {\bibinfo  {journal} {Small Science}\ }\href {https://doi.org/10.1002/smsc.202400161} {10.1002/smsc.202400161} (\bibinfo {year} {2024})\BibitemShut {NoStop}%
\bibitem [{\citenamefont {Pancharatnam}(1956)}]{pancharatnam1956generalized}%
  \BibitemOpen
  \bibfield  {author} {\bibinfo {author} {\bibfnamefont {S.}~\bibnamefont {Pancharatnam}},\ }\bibfield  {title} {\bibinfo {title} {Generalized theory of interference, and its applications: Part i. coherent pencils},\ }in\ \href {https://doi.org/10.1007/BF03046050} {\emph {\bibinfo {booktitle} {Proceedings of the Indian Academy of Sciences-Section A}}},\ Vol.~\bibinfo {volume} {44}\ (\bibinfo {organization} {Springer},\ \bibinfo {year} {1956})\ pp.\ \bibinfo {pages} {247--262}\BibitemShut {NoStop}%
\bibitem [{\citenamefont {Berry}(1984)}]{berry1984quantal}%
  \BibitemOpen
  \bibfield  {author} {\bibinfo {author} {\bibfnamefont {M.~V.}\ \bibnamefont {Berry}},\ }\bibfield  {title} {\bibinfo {title} {Quantal phase factors accompanying adiabatic changes},\ }\href {https://doi.org/10.1098/rspa.1984.0023} {\bibfield  {journal} {\bibinfo  {journal} {Proceedings of the Royal Society of London. A. Mathematical and Physical Sciences}\ }\textbf {\bibinfo {volume} {392}},\ \bibinfo {pages} {45} (\bibinfo {year} {1984})}\BibitemShut {NoStop}%
\bibitem [{\citenamefont {Peskin}(2018)}]{peskin2018introduction}%
  \BibitemOpen
  \bibfield  {author} {\bibinfo {author} {\bibfnamefont {M.~E.}\ \bibnamefont {Peskin}},\ }\href {https://doi.org/10.1201/9780429503559} {\emph {\bibinfo {title} {An Introduction to quantum field theory}}}\ (\bibinfo  {publisher} {CRC press},\ \bibinfo {year} {2018})\BibitemShut {NoStop}%
\bibitem [{\citenamefont {Pokorski}(2000)}]{pokorski2000gauge}%
  \BibitemOpen
  \bibfield  {author} {\bibinfo {author} {\bibfnamefont {S.}~\bibnamefont {Pokorski}},\ }\href {https://doi.org/10.1017/CBO9780511612343} {\emph {\bibinfo {title} {Gauge field theories}}},\ Vol.\ \bibinfo {volume} {136}\ (\bibinfo  {publisher} {Cambridge University Press},\ \bibinfo {year} {2000})\BibitemShut {NoStop}%
\bibitem [{\citenamefont {Mills}(1989)}]{mills1989gauge}%
  \BibitemOpen
  \bibfield  {author} {\bibinfo {author} {\bibfnamefont {R.}~\bibnamefont {Mills}},\ }\bibfield  {title} {\bibinfo {title} {Gauge fields},\ }\href {https://doi.org/10.1119/1.15984} {\bibfield  {journal} {\bibinfo  {journal} {American Journal of Physics}\ }\textbf {\bibinfo {volume} {57}},\ \bibinfo {pages} {493} (\bibinfo {year} {1989})}\BibitemShut {NoStop}%
\bibitem [{\citenamefont {Ivanenko}\ and\ \citenamefont {Sardanashvily}(1983)}]{ivanenko1983gauge}%
  \BibitemOpen
  \bibfield  {author} {\bibinfo {author} {\bibfnamefont {D.}~\bibnamefont {Ivanenko}}\ and\ \bibinfo {author} {\bibfnamefont {G.}~\bibnamefont {Sardanashvily}},\ }\bibfield  {title} {\bibinfo {title} {The gauge treatment of gravity},\ }\href {https://doi.org/10.1016/0370-1573(83)90046-7} {\bibfield  {journal} {\bibinfo  {journal} {Physics Reports}\ }\textbf {\bibinfo {volume} {94}},\ \bibinfo {pages} {1} (\bibinfo {year} {1983})}\BibitemShut {NoStop}%
\bibitem [{\citenamefont {Ryder}(2009)}]{ryder2009introduction}%
  \BibitemOpen
  \bibfield  {author} {\bibinfo {author} {\bibfnamefont {L.}~\bibnamefont {Ryder}},\ }\href {https://doi.org/10.1017/CBO9780511809033} {\emph {\bibinfo {title} {Introduction to general relativity}}}\ (\bibinfo  {publisher} {Cambridge University Press},\ \bibinfo {year} {2009})\BibitemShut {NoStop}%
\bibitem [{\citenamefont {O'Raifeartaigh}(1997)}]{o1997dawning}%
  \BibitemOpen
  \bibfield  {author} {\bibinfo {author} {\bibfnamefont {L.}~\bibnamefont {O'Raifeartaigh}},\ }\href {https://doi.org/10.2307/j.ctv10vm2qt} {\emph {\bibinfo {title} {The dawning of gauge theory}}},\ Vol.\ \bibinfo {volume} {106}\ (\bibinfo  {publisher} {Princeton University Press},\ \bibinfo {year} {1997})\BibitemShut {NoStop}%
\bibitem [{\citenamefont {O’Raifeartaigh}\ and\ \citenamefont {Straumann}(2000)}]{o2000gauge}%
  \BibitemOpen
  \bibfield  {author} {\bibinfo {author} {\bibfnamefont {L.}~\bibnamefont {O’Raifeartaigh}}\ and\ \bibinfo {author} {\bibfnamefont {N.}~\bibnamefont {Straumann}},\ }\bibfield  {title} {\bibinfo {title} {Gauge theory: Historical origins and some modern developments},\ }\href {https://doi.org/10.1103/RevModPhys.72.1} {\bibfield  {journal} {\bibinfo  {journal} {Reviews of Modern Physics}\ }\textbf {\bibinfo {volume} {72}},\ \bibinfo {pages} {1} (\bibinfo {year} {2000})}\BibitemShut {NoStop}%
\bibitem [{\citenamefont {Fujita}\ \emph {et~al.}(2011)\citenamefont {Fujita}, \citenamefont {Jalil}, \citenamefont {Tan},\ and\ \citenamefont {Murakami}}]{fujita2011gauge}%
  \BibitemOpen
  \bibfield  {author} {\bibinfo {author} {\bibfnamefont {T.}~\bibnamefont {Fujita}}, \bibinfo {author} {\bibfnamefont {M.}~\bibnamefont {Jalil}}, \bibinfo {author} {\bibfnamefont {S.}~\bibnamefont {Tan}},\ and\ \bibinfo {author} {\bibfnamefont {S.}~\bibnamefont {Murakami}},\ }\bibfield  {title} {\bibinfo {title} {Gauge fields in spintronics},\ }\bibfield  {journal} {\bibinfo  {journal} {Journal of applied physics}\ }\textbf {\bibinfo {volume} {110}},\ \href {https://doi.org/10.1063/1.3665219} {10.1063/1.3665219} (\bibinfo {year} {2011})\BibitemShut {NoStop}%
\bibitem [{\citenamefont {Palumbo}\ and\ \citenamefont {Goldman}(2019)}]{palumbo2019tensor}%
  \BibitemOpen
  \bibfield  {author} {\bibinfo {author} {\bibfnamefont {G.}~\bibnamefont {Palumbo}}\ and\ \bibinfo {author} {\bibfnamefont {N.}~\bibnamefont {Goldman}},\ }\bibfield  {title} {\bibinfo {title} {Tensor berry connections and their topological invariants},\ }\href {https://doi.org/10.1103/PhysRevB.99.045154} {\bibfield  {journal} {\bibinfo  {journal} {Physical Review B}\ }\textbf {\bibinfo {volume} {99}},\ \bibinfo {pages} {045154} (\bibinfo {year} {2019})}\BibitemShut {NoStop}%
\bibitem [{\citenamefont {Hasegawa}\ and\ \citenamefont {Bollinger}(2005)}]{hasegawa2005rotating}%
  \BibitemOpen
  \bibfield  {author} {\bibinfo {author} {\bibfnamefont {T.}~\bibnamefont {Hasegawa}}\ and\ \bibinfo {author} {\bibfnamefont {J.~J.}\ \bibnamefont {Bollinger}},\ }\bibfield  {title} {\bibinfo {title} {Rotating-radio-frequency ion traps},\ }\href {https://doi.org/10.1103/PhysRevA.72.043403} {\bibfield  {journal} {\bibinfo  {journal} {Physical Review A—Atomic, Molecular, and Optical Physics}\ }\textbf {\bibinfo {volume} {72}},\ \bibinfo {pages} {043403} (\bibinfo {year} {2005})}\BibitemShut {NoStop}%
\bibitem [{\citenamefont {Hughes}\ \emph {et~al.}(2011)\citenamefont {Hughes}, \citenamefont {Lekitsch}, \citenamefont {Broersma},\ and\ \citenamefont {Hensinger}}]{hughes2011microfabricated}%
  \BibitemOpen
  \bibfield  {author} {\bibinfo {author} {\bibfnamefont {M.~D.}\ \bibnamefont {Hughes}}, \bibinfo {author} {\bibfnamefont {B.}~\bibnamefont {Lekitsch}}, \bibinfo {author} {\bibfnamefont {J.~A.}\ \bibnamefont {Broersma}},\ and\ \bibinfo {author} {\bibfnamefont {W.~K.}\ \bibnamefont {Hensinger}},\ }\bibfield  {title} {\bibinfo {title} {Microfabricated ion traps},\ }\href {https://doi.org/10.1080/00107514.2011.601918} {\bibfield  {journal} {\bibinfo  {journal} {Contemporary Physics}\ }\textbf {\bibinfo {volume} {52}},\ \bibinfo {pages} {505} (\bibinfo {year} {2011})}\BibitemShut {NoStop}%
\bibitem [{\citenamefont {Alighanbari}\ \emph {et~al.}(2018)\citenamefont {Alighanbari}, \citenamefont {Hansen}, \citenamefont {Korobov},\ and\ \citenamefont {Schiller}}]{alighanbari2018rotational}%
  \BibitemOpen
  \bibfield  {author} {\bibinfo {author} {\bibfnamefont {S.}~\bibnamefont {Alighanbari}}, \bibinfo {author} {\bibfnamefont {M.~G.}\ \bibnamefont {Hansen}}, \bibinfo {author} {\bibfnamefont {V.}~\bibnamefont {Korobov}},\ and\ \bibinfo {author} {\bibfnamefont {S.}~\bibnamefont {Schiller}},\ }\bibfield  {title} {\bibinfo {title} {Rotational spectroscopy of cold and trapped molecular ions in the lamb--dicke regime},\ }\href {https://doi.org/10.1038/s41567-018-0074-3} {\bibfield  {journal} {\bibinfo  {journal} {Nature Physics}\ }\textbf {\bibinfo {volume} {14}},\ \bibinfo {pages} {555} (\bibinfo {year} {2018})}\BibitemShut {NoStop}%
\bibitem [{\citenamefont {Mudd}\ \emph {et~al.}(2018)\citenamefont {Mudd}, \citenamefont {Martini}, \citenamefont {Zu}, \citenamefont {Kochanek}, \citenamefont {Peterson}, \citenamefont {Kessler}, \citenamefont {Davis}, \citenamefont {Hoormann}, \citenamefont {King}, \citenamefont {Lidman} \emph {et~al.}}]{mudd2018quasar}%
  \BibitemOpen
  \bibfield  {author} {\bibinfo {author} {\bibfnamefont {D.}~\bibnamefont {Mudd}}, \bibinfo {author} {\bibfnamefont {P.}~\bibnamefont {Martini}}, \bibinfo {author} {\bibfnamefont {Y.}~\bibnamefont {Zu}}, \bibinfo {author} {\bibfnamefont {C.}~\bibnamefont {Kochanek}}, \bibinfo {author} {\bibfnamefont {B.~M.}\ \bibnamefont {Peterson}}, \bibinfo {author} {\bibfnamefont {R.}~\bibnamefont {Kessler}}, \bibinfo {author} {\bibfnamefont {T.}~\bibnamefont {Davis}}, \bibinfo {author} {\bibfnamefont {J.}~\bibnamefont {Hoormann}}, \bibinfo {author} {\bibfnamefont {A.}~\bibnamefont {King}}, \bibinfo {author} {\bibfnamefont {C.}~\bibnamefont {Lidman}}, \emph {et~al.},\ }\bibfield  {title} {\bibinfo {title} {Quasar accretion disk sizes from continuum reverberation mapping from the dark energy survey},\ }\href {https://doi.org/10.3847/1538-4357/aac9bb} {\bibfield  {journal} {\bibinfo  {journal} {The Astrophysical Journal}\ }\textbf {\bibinfo {volume} {862}},\ \bibinfo {pages} {123} (\bibinfo {year} {2018})}\BibitemShut
  {NoStop}%
\bibitem [{\citenamefont {Hobbs}\ \emph {et~al.}(2011)\citenamefont {Hobbs}, \citenamefont {Miller}, \citenamefont {Manchester}, \citenamefont {Dempsey}, \citenamefont {Chapman}, \citenamefont {Khoo}, \citenamefont {Applegate}, \citenamefont {Bailes}, \citenamefont {Bhat}, \citenamefont {Bridle} \emph {et~al.}}]{hobbs2011parkes}%
  \BibitemOpen
  \bibfield  {author} {\bibinfo {author} {\bibfnamefont {G.}~\bibnamefont {Hobbs}}, \bibinfo {author} {\bibfnamefont {D.}~\bibnamefont {Miller}}, \bibinfo {author} {\bibfnamefont {R.}~\bibnamefont {Manchester}}, \bibinfo {author} {\bibfnamefont {J.}~\bibnamefont {Dempsey}}, \bibinfo {author} {\bibfnamefont {J.~M.}\ \bibnamefont {Chapman}}, \bibinfo {author} {\bibfnamefont {J.}~\bibnamefont {Khoo}}, \bibinfo {author} {\bibfnamefont {J.}~\bibnamefont {Applegate}}, \bibinfo {author} {\bibfnamefont {M.}~\bibnamefont {Bailes}}, \bibinfo {author} {\bibfnamefont {N.}~\bibnamefont {Bhat}}, \bibinfo {author} {\bibfnamefont {R.}~\bibnamefont {Bridle}}, \emph {et~al.},\ }\bibfield  {title} {\bibinfo {title} {The parkes observatory pulsar data archive},\ }\href {https://doi.org/10.1071/AS11016} {\bibfield  {journal} {\bibinfo  {journal} {Publications of the Astronomical Society of Australia}\ }\textbf {\bibinfo {volume} {28}},\ \bibinfo {pages} {202} (\bibinfo {year} {2011})}\BibitemShut {NoStop}%
\bibitem [{\citenamefont {Dokuchaev}(2014)}]{dokuchaev2014spin}%
  \BibitemOpen
  \bibfield  {author} {\bibinfo {author} {\bibfnamefont {V.~I.}\ \bibnamefont {Dokuchaev}},\ }\bibfield  {title} {\bibinfo {title} {Spin and mass of the nearest supermassive black hole},\ }\href {https://doi.org/10.1007/s10714-014-1832-x} {\bibfield  {journal} {\bibinfo  {journal} {General Relativity and Gravitation}\ }\textbf {\bibinfo {volume} {46}},\ \bibinfo {pages} {1832} (\bibinfo {year} {2014})}\BibitemShut {NoStop}%
\bibitem [{\citenamefont {Melia}\ and\ \citenamefont {Falcke}(2001)}]{melia2001supermassive}%
  \BibitemOpen
  \bibfield  {author} {\bibinfo {author} {\bibfnamefont {F.}~\bibnamefont {Melia}}\ and\ \bibinfo {author} {\bibfnamefont {H.}~\bibnamefont {Falcke}},\ }\bibfield  {title} {\bibinfo {title} {The supermassive black hole at the galactic center},\ }\href {https://doi.org/10.1146/annurev.astro.39.1.309} {\bibfield  {journal} {\bibinfo  {journal} {Annual Review of Astronomy and Astrophysics}\ }\textbf {\bibinfo {volume} {39}},\ \bibinfo {pages} {309} (\bibinfo {year} {2001})}\BibitemShut {NoStop}%
\bibitem [{\citenamefont {Gonzalez-Nelson}\ \emph {et~al.}(2019)\citenamefont {Gonzalez-Nelson}, \citenamefont {Coudert},\ and\ \citenamefont {van Der~Veen}}]{gonzalez2019rotational}%
  \BibitemOpen
  \bibfield  {author} {\bibinfo {author} {\bibfnamefont {A.}~\bibnamefont {Gonzalez-Nelson}}, \bibinfo {author} {\bibfnamefont {F.-X.}\ \bibnamefont {Coudert}},\ and\ \bibinfo {author} {\bibfnamefont {M.~A.}\ \bibnamefont {van Der~Veen}},\ }\bibfield  {title} {\bibinfo {title} {Rotational dynamics of linkers in metal--organic frameworks},\ }\href {https://doi.org/10.3390/nano9030330} {\bibfield  {journal} {\bibinfo  {journal} {Nanomaterials}\ }\textbf {\bibinfo {volume} {9}},\ \bibinfo {pages} {330} (\bibinfo {year} {2019})}\BibitemShut {NoStop}%
\bibitem [{\citenamefont {Jiang}\ \emph {et~al.}(2016)\citenamefont {Jiang}, \citenamefont {Duan}, \citenamefont {Khan},\ and\ \citenamefont {Garcia-Garibay}}]{jiang2016diffusion}%
  \BibitemOpen
  \bibfield  {author} {\bibinfo {author} {\bibfnamefont {X.}~\bibnamefont {Jiang}}, \bibinfo {author} {\bibfnamefont {H.-B.}\ \bibnamefont {Duan}}, \bibinfo {author} {\bibfnamefont {S.~I.}\ \bibnamefont {Khan}},\ and\ \bibinfo {author} {\bibfnamefont {M.~A.}\ \bibnamefont {Garcia-Garibay}},\ }\bibfield  {title} {\bibinfo {title} {Diffusion-controlled rotation of triptycene in a metal--organic framework (mof) sheds light on the viscosity of mof-confined solvent},\ }\href {https://doi.org/10.1021/acscentsci.6b00168} {\bibfield  {journal} {\bibinfo  {journal} {ACS Central Science}\ }\textbf {\bibinfo {volume} {2}},\ \bibinfo {pages} {608} (\bibinfo {year} {2016})}\BibitemShut {NoStop}%
\bibitem [{\citenamefont {Lovas}\ \emph {et~al.}(2005)\citenamefont {Lovas}, \citenamefont {Tiemann}, \citenamefont {Coursey}, \citenamefont {Kotochigova}, \citenamefont {Chang}, \citenamefont {Olsen},\ and\ \citenamefont {Dragoset}}]{lovas2005diatomic}%
  \BibitemOpen
  \bibfield  {author} {\bibinfo {author} {\bibfnamefont {F.~J.}\ \bibnamefont {Lovas}}, \bibinfo {author} {\bibfnamefont {E.}~\bibnamefont {Tiemann}}, \bibinfo {author} {\bibfnamefont {J.~S.}\ \bibnamefont {Coursey}}, \bibinfo {author} {\bibfnamefont {S.~A.}\ \bibnamefont {Kotochigova}}, \bibinfo {author} {\bibfnamefont {J.}~\bibnamefont {Chang}}, \bibinfo {author} {\bibfnamefont {K.}~\bibnamefont {Olsen}},\ and\ \bibinfo {author} {\bibfnamefont {R.~A.}\ \bibnamefont {Dragoset}},\ }\href {https://doi.org/10.18434/T4T59X} {\bibinfo {title} {Diatomic spectral database (version 2.1)}},\ \bibinfo {howpublished} {Online} (\bibinfo {year} {2005}),\ \bibinfo {note} {[Accessed: 2025-10-17]}\BibitemShut {NoStop}%
\bibitem [{\citenamefont {Reimann}\ \emph {et~al.}(2018)\citenamefont {Reimann}, \citenamefont {Doderer}, \citenamefont {Hebestreit}, \citenamefont {Diehl}, \citenamefont {Frimmer}, \citenamefont {Windey}, \citenamefont {Tebbenjohanns},\ and\ \citenamefont {Novotny}}]{reimann2018ghz}%
  \BibitemOpen
  \bibfield  {author} {\bibinfo {author} {\bibfnamefont {R.}~\bibnamefont {Reimann}}, \bibinfo {author} {\bibfnamefont {M.}~\bibnamefont {Doderer}}, \bibinfo {author} {\bibfnamefont {E.}~\bibnamefont {Hebestreit}}, \bibinfo {author} {\bibfnamefont {R.}~\bibnamefont {Diehl}}, \bibinfo {author} {\bibfnamefont {M.}~\bibnamefont {Frimmer}}, \bibinfo {author} {\bibfnamefont {D.}~\bibnamefont {Windey}}, \bibinfo {author} {\bibfnamefont {F.}~\bibnamefont {Tebbenjohanns}},\ and\ \bibinfo {author} {\bibfnamefont {L.}~\bibnamefont {Novotny}},\ }\bibfield  {title} {\bibinfo {title} {Ghz rotation of an optically trapped nanoparticle in vacuum},\ }\href {https://doi.org/10.1103/PhysRevLett.121.033602} {\bibfield  {journal} {\bibinfo  {journal} {Physical review letters}\ }\textbf {\bibinfo {volume} {121}},\ \bibinfo {pages} {033602} (\bibinfo {year} {2018})}\BibitemShut {NoStop}%
\bibitem [{\citenamefont {Ahn}\ \emph {et~al.}(2018)\citenamefont {Ahn}, \citenamefont {Xu}, \citenamefont {Bang}, \citenamefont {Deng}, \citenamefont {Hoang}, \citenamefont {Han}, \citenamefont {Ma},\ and\ \citenamefont {Li}}]{ahn2018optically}%
  \BibitemOpen
  \bibfield  {author} {\bibinfo {author} {\bibfnamefont {J.}~\bibnamefont {Ahn}}, \bibinfo {author} {\bibfnamefont {Z.}~\bibnamefont {Xu}}, \bibinfo {author} {\bibfnamefont {J.}~\bibnamefont {Bang}}, \bibinfo {author} {\bibfnamefont {Y.-H.}\ \bibnamefont {Deng}}, \bibinfo {author} {\bibfnamefont {T.~M.}\ \bibnamefont {Hoang}}, \bibinfo {author} {\bibfnamefont {Q.}~\bibnamefont {Han}}, \bibinfo {author} {\bibfnamefont {R.-M.}\ \bibnamefont {Ma}},\ and\ \bibinfo {author} {\bibfnamefont {T.}~\bibnamefont {Li}},\ }\bibfield  {title} {\bibinfo {title} {Optically levitated nanodumbbell torsion balance and ghz nanomechanical rotor},\ }\href {https://doi.org/10.1103/PhysRevLett.121.033603} {\bibfield  {journal} {\bibinfo  {journal} {Physical review letters}\ }\textbf {\bibinfo {volume} {121}},\ \bibinfo {pages} {033603} (\bibinfo {year} {2018})}\BibitemShut {NoStop}%
\bibitem [{\citenamefont {Kuhn}\ \emph {et~al.}(2017{\natexlab{a}})\citenamefont {Kuhn}, \citenamefont {Kosloff}, \citenamefont {Stickler}, \citenamefont {Patolsky}, \citenamefont {Hornberger}, \citenamefont {Arndt},\ and\ \citenamefont {Millen}}]{kuhn2017full}%
  \BibitemOpen
  \bibfield  {author} {\bibinfo {author} {\bibfnamefont {S.}~\bibnamefont {Kuhn}}, \bibinfo {author} {\bibfnamefont {A.}~\bibnamefont {Kosloff}}, \bibinfo {author} {\bibfnamefont {B.~A.}\ \bibnamefont {Stickler}}, \bibinfo {author} {\bibfnamefont {F.}~\bibnamefont {Patolsky}}, \bibinfo {author} {\bibfnamefont {K.}~\bibnamefont {Hornberger}}, \bibinfo {author} {\bibfnamefont {M.}~\bibnamefont {Arndt}},\ and\ \bibinfo {author} {\bibfnamefont {J.}~\bibnamefont {Millen}},\ }\bibfield  {title} {\bibinfo {title} {Full rotational control of levitated silicon nanorods},\ }\href {https://doi.org/10.1364/OPTICA.4.000356} {\bibfield  {journal} {\bibinfo  {journal} {Optica}\ }\textbf {\bibinfo {volume} {4}},\ \bibinfo {pages} {356} (\bibinfo {year} {2017}{\natexlab{a}})}\BibitemShut {NoStop}%
\bibitem [{\citenamefont {Kuhn}\ \emph {et~al.}(2017{\natexlab{b}})\citenamefont {Kuhn}, \citenamefont {Stickler}, \citenamefont {Kosloff}, \citenamefont {Patolsky}, \citenamefont {Hornberger}, \citenamefont {Arndt},\ and\ \citenamefont {Millen}}]{kuhn2017optically}%
  \BibitemOpen
  \bibfield  {author} {\bibinfo {author} {\bibfnamefont {S.}~\bibnamefont {Kuhn}}, \bibinfo {author} {\bibfnamefont {B.~A.}\ \bibnamefont {Stickler}}, \bibinfo {author} {\bibfnamefont {A.}~\bibnamefont {Kosloff}}, \bibinfo {author} {\bibfnamefont {F.}~\bibnamefont {Patolsky}}, \bibinfo {author} {\bibfnamefont {K.}~\bibnamefont {Hornberger}}, \bibinfo {author} {\bibfnamefont {M.}~\bibnamefont {Arndt}},\ and\ \bibinfo {author} {\bibfnamefont {J.}~\bibnamefont {Millen}},\ }\bibfield  {title} {\bibinfo {title} {Optically driven ultra-stable nanomechanical rotor},\ }\href {https://doi.org/10.1038/s41467-017-01902-9} {\bibfield  {journal} {\bibinfo  {journal} {Nature communications}\ }\textbf {\bibinfo {volume} {8}},\ \bibinfo {pages} {1670} (\bibinfo {year} {2017}{\natexlab{b}})}\BibitemShut {NoStop}%
\bibitem [{\citenamefont {Jin}\ \emph {et~al.}(2021)\citenamefont {Jin}, \citenamefont {Yan}, \citenamefont {Rahman}, \citenamefont {Li}, \citenamefont {Yu},\ and\ \citenamefont {Zhang}}]{jin20216}%
  \BibitemOpen
  \bibfield  {author} {\bibinfo {author} {\bibfnamefont {Y.}~\bibnamefont {Jin}}, \bibinfo {author} {\bibfnamefont {J.}~\bibnamefont {Yan}}, \bibinfo {author} {\bibfnamefont {S.~J.}\ \bibnamefont {Rahman}}, \bibinfo {author} {\bibfnamefont {J.}~\bibnamefont {Li}}, \bibinfo {author} {\bibfnamefont {X.}~\bibnamefont {Yu}},\ and\ \bibinfo {author} {\bibfnamefont {J.}~\bibnamefont {Zhang}},\ }\bibfield  {title} {\bibinfo {title} {6 ghz hyperfast rotation of an optically levitated nanoparticle in vacuum},\ }\href {https://doi.org/10.1364/PRJ.422975} {\bibfield  {journal} {\bibinfo  {journal} {Photonics Research}\ }\textbf {\bibinfo {volume} {9}},\ \bibinfo {pages} {1344} (\bibinfo {year} {2021})}\BibitemShut {NoStop}%
\bibitem [{\citenamefont {Rapp}(1967)}]{rapp1967equatorial}%
  \BibitemOpen
  \bibfield  {author} {\bibinfo {author} {\bibfnamefont {R.~H.}\ \bibnamefont {Rapp}},\ }\bibfield  {title} {\bibinfo {title} {The equatorial radius of the earth and the zero-order undulation of the geoid},\ }\href {https://doi.org/10.1029/JZ072i002p00589} {\bibfield  {journal} {\bibinfo  {journal} {Journal of Geophysical Research}\ }\textbf {\bibinfo {volume} {72}},\ \bibinfo {pages} {589} (\bibinfo {year} {1967})}\BibitemShut {NoStop}%
\bibitem [{\citenamefont {Demtr{\"o}der}(2024)}]{demtroder2024astrophysics}%
  \BibitemOpen
  \bibfield  {author} {\bibinfo {author} {\bibfnamefont {W.}~\bibnamefont {Demtr{\"o}der}},\ }\href {https://doi.org/10.1007/978-3-031-22135-4} {\emph {\bibinfo {title} {Astrophysics}}}\ (\bibinfo  {publisher} {Springer Nature},\ \bibinfo {year} {2024})\BibitemShut {NoStop}%
\bibitem [{\citenamefont {Hessels}\ \emph {et~al.}(2006)\citenamefont {Hessels}, \citenamefont {Ransom}, \citenamefont {Stairs}, \citenamefont {Freire}, \citenamefont {Kaspi},\ and\ \citenamefont {Camilo}}]{hessels2006radio}%
  \BibitemOpen
  \bibfield  {author} {\bibinfo {author} {\bibfnamefont {J.~W.}\ \bibnamefont {Hessels}}, \bibinfo {author} {\bibfnamefont {S.~M.}\ \bibnamefont {Ransom}}, \bibinfo {author} {\bibfnamefont {I.~H.}\ \bibnamefont {Stairs}}, \bibinfo {author} {\bibfnamefont {P.~C.}\ \bibnamefont {Freire}}, \bibinfo {author} {\bibfnamefont {V.~M.}\ \bibnamefont {Kaspi}},\ and\ \bibinfo {author} {\bibfnamefont {F.}~\bibnamefont {Camilo}},\ }\bibfield  {title} {\bibinfo {title} {A radio pulsar spinning at 716 hz},\ }\href {https://doi.org/10.1126/science.1123430} {\bibfield  {journal} {\bibinfo  {journal} {Science}\ }\textbf {\bibinfo {volume} {311}},\ \bibinfo {pages} {1901} (\bibinfo {year} {2006})}\BibitemShut {NoStop}%
\bibitem [{\citenamefont {Shakura}\ and\ \citenamefont {Sunyaev}(1973)}]{shakura1973black}%
  \BibitemOpen
  \bibfield  {author} {\bibinfo {author} {\bibfnamefont {N.~I.}\ \bibnamefont {Shakura}}\ and\ \bibinfo {author} {\bibfnamefont {R.~A.}\ \bibnamefont {Sunyaev}},\ }\bibfield  {title} {\bibinfo {title} {Black holes in binary systems. observational appearance.},\ }\href {https://doi.org/10.1017/S007418090010035X} {\bibfield  {journal} {\bibinfo  {journal} {Astronomy and Astrophysics, Vol. 24, p. 337-355}\ }\textbf {\bibinfo {volume} {24}},\ \bibinfo {pages} {337} (\bibinfo {year} {1973})}\BibitemShut {NoStop}%
\bibitem [{\citenamefont {Morgan}\ \emph {et~al.}(2010)\citenamefont {Morgan}, \citenamefont {Kochanek}, \citenamefont {Morgan},\ and\ \citenamefont {Falco}}]{morgan2010quasar}%
  \BibitemOpen
  \bibfield  {author} {\bibinfo {author} {\bibfnamefont {C.~W.}\ \bibnamefont {Morgan}}, \bibinfo {author} {\bibfnamefont {C.}~\bibnamefont {Kochanek}}, \bibinfo {author} {\bibfnamefont {N.~D.}\ \bibnamefont {Morgan}},\ and\ \bibinfo {author} {\bibfnamefont {E.~E.}\ \bibnamefont {Falco}},\ }\bibfield  {title} {\bibinfo {title} {The quasar accretion disk size--black hole mass relation},\ }\href {https://doi.org/10.1088/0004-637X/712/2/1129} {\bibfield  {journal} {\bibinfo  {journal} {The Astrophysical Journal}\ }\textbf {\bibinfo {volume} {712}},\ \bibinfo {pages} {1129} (\bibinfo {year} {2010})}\BibitemShut {NoStop}%
\bibitem [{\citenamefont {Bentz}\ and\ \citenamefont {Katz}(2015)}]{bentz2015agn}%
  \BibitemOpen
  \bibfield  {author} {\bibinfo {author} {\bibfnamefont {M.~C.}\ \bibnamefont {Bentz}}\ and\ \bibinfo {author} {\bibfnamefont {S.}~\bibnamefont {Katz}},\ }\bibfield  {title} {\bibinfo {title} {The agn black hole mass database},\ }\href {https://doi.org/10.1086/679601} {\bibfield  {journal} {\bibinfo  {journal} {Publications of the Astronomical Society of the Pacific}\ }\textbf {\bibinfo {volume} {127}},\ \bibinfo {pages} {67} (\bibinfo {year} {2015})}\BibitemShut {NoStop}%
\bibitem [{\citenamefont {Bjorken}\ \emph {et~al.}(1965)\citenamefont {Bjorken}, \citenamefont {Drell},\ and\ \citenamefont {Mansfield}}]{bjorken1965relativistic}%
  \BibitemOpen
  \bibfield  {author} {\bibinfo {author} {\bibfnamefont {J.~D.}\ \bibnamefont {Bjorken}}, \bibinfo {author} {\bibfnamefont {S.~D.}\ \bibnamefont {Drell}},\ and\ \bibinfo {author} {\bibfnamefont {J.}~\bibnamefont {Mansfield}},\ }\href {https://doi.org/10.1063/1.3047288} {\bibinfo {title} {Relativistic quantum mechanics}} (\bibinfo {year} {1965})\BibitemShut {NoStop}%
\bibitem [{\citenamefont {Gergely}(2002)}]{gergely2002hamiltonian}%
  \BibitemOpen
  \bibfield  {author} {\bibinfo {author} {\bibfnamefont {L.~{\'A}.}\ \bibnamefont {Gergely}},\ }\bibfield  {title} {\bibinfo {title} {On hamiltonian formulations of the schr{\"o}dinger system},\ }\href {https://doi.org/10.1006/aphy.2002.6262} {\bibfield  {journal} {\bibinfo  {journal} {Annals of Physics}\ }\textbf {\bibinfo {volume} {298}},\ \bibinfo {pages} {394} (\bibinfo {year} {2002})}\BibitemShut {NoStop}%
\bibitem [{\citenamefont {Griffiths}(2020)}]{griffiths2020introduction}%
  \BibitemOpen
  \bibfield  {author} {\bibinfo {author} {\bibfnamefont {D.}~\bibnamefont {Griffiths}},\ }\href {https://doi.org/10.1002/9783527618460} {\emph {\bibinfo {title} {Introduction to elementary particles}}}\ (\bibinfo  {publisher} {John Wiley \& Sons},\ \bibinfo {year} {2020})\BibitemShut {NoStop}%
\bibitem [{\citenamefont {Gangaraj}\ \emph {et~al.}(2017)\citenamefont {Gangaraj}, \citenamefont {Silveirinha},\ and\ \citenamefont {Hanson}}]{gangaraj2017berry}%
  \BibitemOpen
  \bibfield  {author} {\bibinfo {author} {\bibfnamefont {S.~A.~H.}\ \bibnamefont {Gangaraj}}, \bibinfo {author} {\bibfnamefont {M.~G.}\ \bibnamefont {Silveirinha}},\ and\ \bibinfo {author} {\bibfnamefont {G.~W.}\ \bibnamefont {Hanson}},\ }\bibfield  {title} {\bibinfo {title} {Berry phase, berry connection, and chern number for a continuum bianisotropic material from a classical electromagnetics perspective},\ }\href {https://doi.org/0.1109/JMMCT.2017.2654962} {\bibfield  {journal} {\bibinfo  {journal} {IEEE journal on multiscale and multiphysics computational techniques}\ }\textbf {\bibinfo {volume} {2}},\ \bibinfo {pages} {3} (\bibinfo {year} {2017})}\BibitemShut {NoStop}%
\bibitem [{\citenamefont {Magpantay}(1994)}]{magpantay1994coulomb}%
  \BibitemOpen
  \bibfield  {author} {\bibinfo {author} {\bibfnamefont {J.~A.}\ \bibnamefont {Magpantay}},\ }\bibfield  {title} {\bibinfo {title} {The coulomb gauge revisited},\ }\href {https://doi.org/10.1143/ptp/91.3.573} {\bibfield  {journal} {\bibinfo  {journal} {Progress of theoretical physics}\ }\textbf {\bibinfo {volume} {91}},\ \bibinfo {pages} {573} (\bibinfo {year} {1994})}\BibitemShut {NoStop}%
\bibitem [{\citenamefont {Ferrero}\ \emph {et~al.}(2025)\citenamefont {Ferrero}, \citenamefont {Han},\ and\ \citenamefont {Liu}}]{ferrero2025one}%
  \BibitemOpen
  \bibfield  {author} {\bibinfo {author} {\bibfnamefont {R.}~\bibnamefont {Ferrero}}, \bibinfo {author} {\bibfnamefont {M.}~\bibnamefont {Han}},\ and\ \bibinfo {author} {\bibfnamefont {H.}~\bibnamefont {Liu}},\ }\bibfield  {title} {\bibinfo {title} {One-loop effective action from the coherent state path integral of loop quantum gravity},\ }\href {https://doi.org/10.1103/r5gq-k1ss} {\bibfield  {journal} {\bibinfo  {journal} {Physical Review D}\ }\textbf {\bibinfo {volume} {112}},\ \bibinfo {pages} {024033} (\bibinfo {year} {2025})}\BibitemShut {NoStop}%
\bibitem [{\citenamefont {Scharnhorst}(1997)}]{scharnhorst1997functional}%
  \BibitemOpen
  \bibfield  {author} {\bibinfo {author} {\bibfnamefont {K.}~\bibnamefont {Scharnhorst}},\ }\bibfield  {title} {\bibinfo {title} {Functional integral equation for the complete effective action in quantum field theory},\ }\href {https://doi.org/10.1007/BF02435737} {\bibfield  {journal} {\bibinfo  {journal} {International Journal of Theoretical Physics}\ }\textbf {\bibinfo {volume} {36}},\ \bibinfo {pages} {281} (\bibinfo {year} {1997})}\BibitemShut {NoStop}%
\bibitem [{\citenamefont {Zee}(2010)}]{zee2010quantum}%
  \BibitemOpen
  \bibfield  {author} {\bibinfo {author} {\bibfnamefont {A.}~\bibnamefont {Zee}},\ }\href@noop {} {\emph {\bibinfo {title} {Quantum field theory in a nutshell}}},\ Vol.~\bibinfo {volume} {7}\ (\bibinfo  {publisher} {Princeton university press},\ \bibinfo {year} {2010})\BibitemShut {NoStop}%
\bibitem [{\citenamefont {Coon}\ and\ \citenamefont {Holstein}(2002)}]{coon2002anomalies}%
  \BibitemOpen
  \bibfield  {author} {\bibinfo {author} {\bibfnamefont {S.~A.}\ \bibnamefont {Coon}}\ and\ \bibinfo {author} {\bibfnamefont {B.~R.}\ \bibnamefont {Holstein}},\ }\bibfield  {title} {\bibinfo {title} {Anomalies in quantum mechanics: the 1/r 2 potential},\ }\href {https://doi.org/10.1119/1.1456071} {\bibfield  {journal} {\bibinfo  {journal} {American Journal of Physics}\ }\textbf {\bibinfo {volume} {70}},\ \bibinfo {pages} {513} (\bibinfo {year} {2002})}\BibitemShut {NoStop}%
\bibitem [{\citenamefont {Griffiths}(2023)}]{griffiths2023introduction}%
  \BibitemOpen
  \bibfield  {author} {\bibinfo {author} {\bibfnamefont {D.~J.}\ \bibnamefont {Griffiths}},\ }\href@noop {} {\emph {\bibinfo {title} {Introduction to electrodynamics}}}\ (\bibinfo  {publisher} {Cambridge University Press},\ \bibinfo {year} {2023})\BibitemShut {NoStop}%
\bibitem [{\citenamefont {Talin}\ \emph {et~al.}(2008)\citenamefont {Talin}, \citenamefont {Calisti}, \citenamefont {Dufty},\ and\ \citenamefont {Pogorelov}}]{talin2008electron}%
  \BibitemOpen
  \bibfield  {author} {\bibinfo {author} {\bibfnamefont {B.}~\bibnamefont {Talin}}, \bibinfo {author} {\bibfnamefont {A.}~\bibnamefont {Calisti}}, \bibinfo {author} {\bibfnamefont {J.~W.}\ \bibnamefont {Dufty}},\ and\ \bibinfo {author} {\bibfnamefont {I.~V.}\ \bibnamefont {Pogorelov}},\ }\bibfield  {title} {\bibinfo {title} {Electron dynamics at a positive ion},\ }\href {https://doi.org/10.1103/PhysRevE.77.036410} {\bibfield  {journal} {\bibinfo  {journal} {Physical Review E—Statistical, Nonlinear, and Soft Matter Physics}\ }\textbf {\bibinfo {volume} {77}},\ \bibinfo {pages} {036410} (\bibinfo {year} {2008})}\BibitemShut {NoStop}%
\bibitem [{\citenamefont {Decius}(1968)}]{decius1968dipolar}%
  \BibitemOpen
  \bibfield  {author} {\bibinfo {author} {\bibfnamefont {J.}~\bibnamefont {Decius}},\ }\bibfield  {title} {\bibinfo {title} {Dipolar coupling and molecular vibration in crystals. i. general theory},\ }\href {https://doi.org/10.1063/1.1670236} {\bibfield  {journal} {\bibinfo  {journal} {The Journal of Chemical Physics}\ }\textbf {\bibinfo {volume} {49}},\ \bibinfo {pages} {1387} (\bibinfo {year} {1968})}\BibitemShut {NoStop}%
\bibitem [{\citenamefont {Zan}\ \emph {et~al.}(2024)\citenamefont {Zan}, \citenamefont {Guo}, \citenamefont {Deng}, \citenamefont {Huang}, \citenamefont {Liu}, \citenamefont {Wu}, \citenamefont {Yuan}, \citenamefont {Zhao}, \citenamefont {Peng}, \citenamefont {Li} \emph {et~al.}}]{zan2024electron}%
  \BibitemOpen
  \bibfield  {author} {\bibinfo {author} {\bibfnamefont {X.}~\bibnamefont {Zan}}, \bibinfo {author} {\bibfnamefont {X.}~\bibnamefont {Guo}}, \bibinfo {author} {\bibfnamefont {A.}~\bibnamefont {Deng}}, \bibinfo {author} {\bibfnamefont {Z.}~\bibnamefont {Huang}}, \bibinfo {author} {\bibfnamefont {L.}~\bibnamefont {Liu}}, \bibinfo {author} {\bibfnamefont {F.}~\bibnamefont {Wu}}, \bibinfo {author} {\bibfnamefont {Y.}~\bibnamefont {Yuan}}, \bibinfo {author} {\bibfnamefont {J.}~\bibnamefont {Zhao}}, \bibinfo {author} {\bibfnamefont {Y.}~\bibnamefont {Peng}}, \bibinfo {author} {\bibfnamefont {L.}~\bibnamefont {Li}}, \emph {et~al.},\ }\bibfield  {title} {\bibinfo {title} {Electron/infrared-phonon coupling in abc trilayer graphene},\ }\href {https://doi.org/10.1038/s41467-024-46129-7} {\bibfield  {journal} {\bibinfo  {journal} {Nature Communications}\ }\textbf {\bibinfo {volume} {15}},\ \bibinfo {pages} {1888} (\bibinfo {year} {2024})}\BibitemShut {NoStop}%
\bibitem [{\citenamefont {Hehl}\ \emph {et~al.}(1991)\citenamefont {Hehl}, \citenamefont {Lemke},\ and\ \citenamefont {Mielke}}]{hehl1991two}%
  \BibitemOpen
  \bibfield  {author} {\bibinfo {author} {\bibfnamefont {F.~W.}\ \bibnamefont {Hehl}}, \bibinfo {author} {\bibfnamefont {J.}~\bibnamefont {Lemke}},\ and\ \bibinfo {author} {\bibfnamefont {E.~W.}\ \bibnamefont {Mielke}},\ }\bibfield  {title} {\bibinfo {title} {Two lectures on fermions and gravity},\ }in\ \href {https://doi.org/10.1007/978-3-642-76353-3_3} {\emph {\bibinfo {booktitle} {Geometry and Theoretical Physics}}}\ (\bibinfo  {publisher} {Springer},\ \bibinfo {year} {1991})\ pp.\ \bibinfo {pages} {56--140}\BibitemShut {NoStop}%
\bibitem [{\citenamefont {Ni}(1977)}]{ni1977proper}%
  \BibitemOpen
  \bibfield  {author} {\bibinfo {author} {\bibfnamefont {W.-T.}\ \bibnamefont {Ni}},\ }\bibfield  {title} {\bibinfo {title} {On the proper reference frame and local coordinates of an accelerated observer in special relativity},\ }\href@noop {} {\bibfield  {journal} {\bibinfo  {journal} {Chinese Journal of Physics}\ }\textbf {\bibinfo {volume} {15}},\ \bibinfo {pages} {51} (\bibinfo {year} {1977})}\BibitemShut {NoStop}%
\bibitem [{\citenamefont {Sciama}(1963)}]{sciama1963recent}%
  \BibitemOpen
  \bibfield  {author} {\bibinfo {author} {\bibfnamefont {D.}~\bibnamefont {Sciama}},\ }\href@noop {} {\bibinfo {title} {Recent developments in general relativity, p. 415}} (\bibinfo {year} {1963})\BibitemShut {NoStop}%
\bibitem [{\citenamefont {Hehl}(1985)}]{hehl1985kinematics}%
  \BibitemOpen
  \bibfield  {author} {\bibinfo {author} {\bibfnamefont {F.~W.}\ \bibnamefont {Hehl}},\ }\bibfield  {title} {\bibinfo {title} {On the kinematics of the torsion of space-time},\ }\href {https://doi.org/10.1007/BF01889281} {\bibfield  {journal} {\bibinfo  {journal} {Foundations of Physics}\ }\textbf {\bibinfo {volume} {15}},\ \bibinfo {pages} {451} (\bibinfo {year} {1985})}\BibitemShut {NoStop}%
\bibitem [{\citenamefont {Kibble}(1961)}]{kibble1961lorentz}%
  \BibitemOpen
  \bibfield  {author} {\bibinfo {author} {\bibfnamefont {T.~W.}\ \bibnamefont {Kibble}},\ }\bibfield  {title} {\bibinfo {title} {Lorentz invariance and the gravitational field},\ }\href {https://doi.org/10.1063/1.1703702} {\bibfield  {journal} {\bibinfo  {journal} {Journal of mathematical physics}\ }\textbf {\bibinfo {volume} {2}},\ \bibinfo {pages} {212} (\bibinfo {year} {1961})}\BibitemShut {NoStop}%
\bibitem [{\citenamefont {Xiao}\ \emph {et~al.}(2009)\citenamefont {Xiao}, \citenamefont {Shi}, \citenamefont {Clougherty},\ and\ \citenamefont {Niu}}]{xiao2009polarization}%
  \BibitemOpen
  \bibfield  {author} {\bibinfo {author} {\bibfnamefont {D.}~\bibnamefont {Xiao}}, \bibinfo {author} {\bibfnamefont {J.}~\bibnamefont {Shi}}, \bibinfo {author} {\bibfnamefont {D.~P.}\ \bibnamefont {Clougherty}},\ and\ \bibinfo {author} {\bibfnamefont {Q.}~\bibnamefont {Niu}},\ }\bibfield  {title} {\bibinfo {title} {Polarization and adiabatic pumping in inhomogeneous crystals},\ }\href {https://doi.org/10.1103/PhysRevLett.102.087602} {\bibfield  {journal} {\bibinfo  {journal} {Physical review letters}\ }\textbf {\bibinfo {volume} {102}},\ \bibinfo {pages} {087602} (\bibinfo {year} {2009})}\BibitemShut {NoStop}%
\bibitem [{\citenamefont {Xiao}\ \emph {et~al.}(2010)\citenamefont {Xiao}, \citenamefont {Chang},\ and\ \citenamefont {Niu}}]{xiao2010berry}%
  \BibitemOpen
  \bibfield  {author} {\bibinfo {author} {\bibfnamefont {D.}~\bibnamefont {Xiao}}, \bibinfo {author} {\bibfnamefont {M.-C.}\ \bibnamefont {Chang}},\ and\ \bibinfo {author} {\bibfnamefont {Q.}~\bibnamefont {Niu}},\ }\bibfield  {title} {\bibinfo {title} {Berry phase effects on electronic properties},\ }\href {https://doi.org/10.1103/RevModPhys.82.1959} {\bibfield  {journal} {\bibinfo  {journal} {Reviews of modern physics}\ }\textbf {\bibinfo {volume} {82}},\ \bibinfo {pages} {1959} (\bibinfo {year} {2010})}\BibitemShut {NoStop}%
\bibitem [{\citenamefont {Landau}\ \emph {et~al.}(1937)\citenamefont {Landau} \emph {et~al.}}]{landau1937theory}%
  \BibitemOpen
  \bibfield  {author} {\bibinfo {author} {\bibfnamefont {L.~D.}\ \bibnamefont {Landau}} \emph {et~al.},\ }\bibfield  {title} {\bibinfo {title} {On the theory of phase transitions},\ }\href@noop {} {\bibfield  {journal} {\bibinfo  {journal} {Zh. eksp. teor. Fiz}\ }\textbf {\bibinfo {volume} {7}},\ \bibinfo {pages} {926} (\bibinfo {year} {1937})}\BibitemShut {NoStop}%
\bibitem [{\citenamefont {Thonhauser}(2011)}]{thonhauser2011theory}%
  \BibitemOpen
  \bibfield  {author} {\bibinfo {author} {\bibfnamefont {T.}~\bibnamefont {Thonhauser}},\ }\bibfield  {title} {\bibinfo {title} {Theory of orbital magnetization in solids},\ }\href {https://doi.org/10.1142/S0217979211058912} {\bibfield  {journal} {\bibinfo  {journal} {International Journal of Modern Physics B}\ }\textbf {\bibinfo {volume} {25}},\ \bibinfo {pages} {1429} (\bibinfo {year} {2011})}\BibitemShut {NoStop}%
\bibitem [{\citenamefont {Resta}(2010)}]{resta2010electrical}%
  \BibitemOpen
  \bibfield  {author} {\bibinfo {author} {\bibfnamefont {R.}~\bibnamefont {Resta}},\ }\bibfield  {title} {\bibinfo {title} {Electrical polarization and orbital magnetization: the modern theories},\ }\href {https://doi.org/10.1088/0953-8984/22/12/123201} {\bibfield  {journal} {\bibinfo  {journal} {Journal of Physics: Condensed Matter}\ }\textbf {\bibinfo {volume} {22}},\ \bibinfo {pages} {123201} (\bibinfo {year} {2010})}\BibitemShut {NoStop}%
\end{thebibliography}
\end{document}